\documentclass[useAMS,usenatbib]{mn2e}

\usepackage{graphicx}
\usepackage{color}
\usepackage{deluxetable}
\usepackage{xspace}
\usepackage{amsmath}

\usepackage{hyperref}

\usepackage{natbib}
\def\apj{ApJ}
\def\mnras{MNRAS}

\def\araa{ARA\&A}                
\def\aap{A\&A}                   
\def\aj{AJ}                      
\def\apjs{ApJS}                  
\def\apjl{ApJ}                   

\def\nar{New Astron. Rev.}		 
\def\rmxaa{Rev. Mex. Astron. Astrofis.}

\newcommand{\mumeter}{\,$\micron$\xspace}
\newcommand{\rsub}{$r_{sub}$\xspace}
\newcommand{\lx}{$L_X$\xspace}
\newcommand{\lxrsub}{$L_X$\,--\,$r_{sub}$\xspace}
\newcommand{\lxlmir}{$L_X$\,--\,$L_{\mathrm{MIR}}$\xspace}
\newcommand{\pahsurfflux}{$\Sigma_{PAH}$\xspace}

\title[PAH features within few hundred parsecs of AGN]{PAH features within few hundred parsecs of active galactic nuclei\thanks{Based on ESO observing programs 075.B-0163, 078.B-0303, 080.B-0240, 280.B-5068, 082.B-0299, 083.B-0239, 086.B-0919, and 085.B-0251.}}

\author[J. J. Jensen et al.]{J. J. Jensen,$^{1,2}$\thanks{E-mail: juel@dark-cosmology.dk} S. F. H\"{o}nig,$^{2}$ S. Rakshit,$^{3,4,5}$ A. Alonso-Herrero,$^{6}$ D. Asmus,$^{7}$ 
\newauthor
P. Gandhi,$^{2}$ M. Kishimoto,$^{8}$ A. Smette$^{7}$ and K. R. W. Tristram$^{7}$\\
$^{1}$Dark Cosmology Centre, Niels Bohr Institute, University of Copenhagen, Juliane Maries Vej 30, DK-2100 Copenhagen \O, Denmark \\
$^{2}$Department of Physics \& Astronomy, University of Southampton, Southampton SO17 1BJ, UK \\
$^{3}$Laboratoire Lagrange, UMR 7293, University of Nice Sophia-Antipolis, CNRS, France  \\
$^{4}$Observatoire de la Cote D'Azur, BP 4229, F-06304 Nice Cedex 4, France \\
$^{5}$Indian Institute of Astrophysics, Block II, Koramangala, Bangalore-560034, India \\
$^{6}$Centro de Astrobiolog\'{\i}a, CSIC-INTA, ESAC Campus, E-28692
Villanueva de la Ca\~nada, Madrid, Spain \\
$^{7}$European Southern Observatory, Casilla 19001, Santiago 19, Chile \\
$^{8}$Kyoto Sangyo University, Motoyama, Kamigamo, Kita-ku, Kyoto, 603-8555 Japan}
\begin{document}

\date{Accepted 2015 October 20. Received 2015 October 14; in original form 2014 October 11}

\pagerange{\pageref{firstpage}--\pageref{lastpage}} \pubyear{2016}

\maketitle

\label{firstpage}

\begin{abstract}
Spectral features from poly-cyclic aromatic hydrocarbon (PAH) molecules observed in the mid-infrared (mid-IR) range are typically used to infer the amount of recent and ongoing star formation on kiloparsec scales around active galactic nuclei (AGN) where more traditional methods fail. This method assumes that the observed PAH features are excited predominantly by star formation. With current ground-based telescopes and the upcoming JWST, much smaller spatial scales can be probed and we aim at testing if this assumption still holds in the range of few tens to few hundreds of parsecs. For that, we spatially map the emitted 11.3\mumeter PAH surface flux as a function of distance from $0.4-4$~arcsec from the centre in 28 nearby AGN using ground-based high-angular resolution mid-IR spectroscopy. We detect and extract the 11.3\mumeter PAH feature in 13 AGN. The fluxes within each aperture are scaled to a luminosity-normalised distance from the nucleus to be able to compare intrinsic spatial scales of AGN radiation spanning about 2 orders of magnitude in luminosity. For this, we establish an empirical relation between the absorption-corrected X-ray luminosity and the sublimation radius in these sources. Once normalised, the radial profiles of the emitted PAH surface flux show similar radial slopes, with a power-law index of approximately $-1.1$, and similar absolute values, consistent within a factor of a few of each other as expected from the uncertainty in the intrinsic scale estimate. We interpret this as evidence that the profiles are caused by a common compact central physical process, either the AGN itself or circumnuclear star formation linked in strength to the AGN power. A photoionisation-based model of an AGN exciting dense clouds in its environment can reproduce the observed radial slope and confirms that the AGN radiation field is strong enough to explain the observed PAH surface fluxes within $\sim10-500$\,pc of the nucleus. Our results advice caution in the use of PAH emission as a star formation tracer within a kpc around AGN.
\end{abstract}


\section{Introduction}\label{sec:intro}
Understanding galaxy formation and evolution is a fundamental part of current astronomical research. Observed strong correlations between the mass of the central supermassive black holes (SMBHs), believed to be ubiquitous in all galaxies, and the properties of their host galaxy bulge \citep{magorrian1998,wandel1999,ferraresemerritt2000,gebhardt2000b,ferrarese2001,mcluredunlop2001,marconihunt2003} indicates that some kind of feedback or co-evolution is at play. One piece of this puzzle is to understand the connection between the growth of a SMBH during its phase of active accretion (``active galactic nucleus''; AGN) and the star formation in its host galaxy bulge. 

In the standard model of unification, an AGN is fueled by accretion of gas onto a central SMBH \citep[][and references therein]{alexander2012}. Within this scenario, the accretion of gas into the central parts of the galaxy will almost certainly lead to nuclear star formation, as indicated by observations and simulations alike \citep{kawakatu2008,hopkins2010}. Models by \citet{hopkins2012} predict a correlation between the star formation rate and the black hole accretion rate. However, the black holes accretion phase is slightly delayed with respect to the peak of star formation. Observational evidence of AGN activity in post-starburst galaxies \citep{goto2006,wild2010} and star formation in AGN \citep{davies2007,ramosalmeida2013,esquej2014} supports this picture. Other simulations suggest that most star formation occurs in a circumnuclear ring, which could be coincident with the dusty torus of the AGN unification model \citep{wada2002,thompson2005,kawakatu2008}.

Star formation is traditionally probed by UV emission, H$\alpha$, Pa$\alpha$, [Ne\,{\sc ii}] 12.8\mumeter, and modelling of stellar populations. However, in the vicinity of AGN, emission from the accretion process contaminates these tracers. Alternatively, emission features in the mid-infrared (mid-IR) provide a powerful tool to test recent and ongoing star formation --- most prominent the 6.2, 7.7, 8.6, and 11.3\mumeter emission complexes. These aromatic features (hereafter PAH), assumed to originate from stretching and bending modes of poly-cyclic aromatic hydrocarbon molecules \citep[][]{leger1984,allamandola1985,allamandola1989,tielens2008}, are supposedly excited by UV radiation from young massive O- and B-stars \citep{roche1985,roche1991}. The strength of these features has been shown to be a good tracer of star formation rates due to their strong correlation with the infrared luminosity in pure starbursts \citep[e.g.,][]{peeters2004,brandl2006}.

PAH features are surprisingly abundant even in the AGN proximity \citep[e.g.,][]{roche1991,clavel2000,siebenmorgen2004,davies2007,smith2007,deo2009,wu2009,diamondstanic2010,diamondstanic2012,diazsantos2010,honig2010,sales2010,tommasin2010,gonzalezmartin2013,alonsoherrero2014,alonsoherrero2016,esquej2014}, although they can be easily destroyed by strong radiation fields \citep{voit1992}. On the other hand, PAH features are suppressed in AGN compared to star forming galaxies \citep{roche1991,peeters2004,smith2007}, with the caveat that the decrease found in PAH equivalent width can be interpreted as being due to increased dilution from the AGN continuum rather than destruction of the PAH molecules (see for example \citet{esquej2014} and \citet{alonsoherrero2014} for a discussion of this issue). More specifically, \citet{diamondstanic2010} find the 11.3\mumeter feature to be a good tracer of star formation in AGN while the 6.2, 7.7, and 8.6\mumeter features are too suppressed for this purpose. This relatively larger suppression of PAH features at shorter wavelengths was interpreted as selective destruction of the smallest PAH molecules by the radiation field from the AGN \citep{smith2007}. On the contrary, the survival of PAH grains in the nuclear regions of AGN may be related to (self-)shielding from the nuclear radiation field in dense regions \citep[e.g.,][]{esquej2014,alonsoherrero2014}. 

Most of the previous assessments of PAH in the AGN vicinity assume that the PAH features are exclusively excited by circumnuclear kpc-scaled star formation. In this work, we study the spatially resolved 11.3\,\mumeter PAH feature at unprecedented spatial resolution from $\sim$10\,pc to $<$1000\,pc. A specific focus is put on the question if the AGN may play a significant role in exciting the PAH grains on these small, previously untested spatial scales. While it may be difficult to spectrally disentangle both contributions, we follow the idea of spatially mapping the emitted surface flux of the PAH on scales of few tens of parsec around the AGN and assess the spatial distribution as well as the energy budget. For this, we establish the radial distribution of the 11.3\mumeter PAH feature from within $\sim10-1000$ pc around our sample of nearby AGN. We then discuss the observed fluxes, both in physical scaling and AGN-luminosity normalised scales. The results are put into physical context using CLOUDY simulations and a discussion of stellar clusters in two examples. A brief description of the observations and the data is presented in \S~\ref{sec:obsdata} and our method for extracting the 11.3\mumeter PAH feature is described in \S~\ref{sec:method}. In \S~\ref{sec:lxrsub} we present a novel relation between the intrinsic X-ray luminosity of an AGN and its dust sublimation radius to be able to compare the same intrinsic scales for AGN with different luminosities. In \S~\ref{sec:results} we present the radial dependence of the PAH features, and we discuss these results in the context of CLOUDY models and PAH excitation by stars in \S~\ref{sec:cloudy}. Finally, in \S~\ref{sec:conclusion}, we present our conclusions.

\section{Observations and data}\label{sec:obsdata}
The goal of this paper is to spatially map the emitted 11.3\mumeter PAH emission in the vicinity of AGN. For this purpose we use archival low-resolution 8--13\mumeter spectrosopic data from Very Large Telescope (VLT) Spectrometer and Imager for the mid-InfraRed \citep[VISIR;][]{lagage2004} observations of a sample of local AGN \citep{honig2010,burtscher2013}. We supplement these data with new, unpublished observations taken in 2005, 2008, and 2010, which represent VISIR spectra for Circinus, ESO~138-G001, F9, F49, and NGC 1068. Information on these previously unpublished observations are shown in Table~\ref{tab:newdata}. For the archival data, a slit width of 0.8~arcsec was used, except for NGC 4507, MCG-3-34-64 and IC5063 observed with a 1.0~arcsec slit, and ESO 323-G77, MCG-6-30-15, IC4329A, NGC 5643 and NGC 5995 observed with a 0.75~arcsec slit. More information on the archival data can be found in the respective sources listed in Table~\ref{tab:agnsample}. The sample represents the highest spatial resolution observations available in the mid-IR and is typical of the local population of AGN. The data reduction and calibration follows the methods and tools outlined in \citet{honig2010}. This includes the standard ESO pipeline \footnote{http://www.eso.org/sci/software/pipelines/visir/visir-pipe-recipes.html} and additional procedures to model and remove a periodic background pattern from the observations. Moreover, flux calibration is carried out on the full 2-dimensional spectrum instead of only an optimised aperture. These ground-based observations achieve angular resolutions of 0.3--0.4~arcsec. 

The final sample contains 28 nearby AGN at distances from 4.2 to 248 Mpc (median distance of 45.3 Mpc; see Table~\ref{tab:agnsample}). We supplement these data with absorption-corrected 2--10~keV luminosities compiled by \citet{asmus2015}. These data allow us to estimate the intrinsic strength of the AGN and establish a luminosity independent ``intrinsic'' scaling for the AGN (see Sect.~\ref{sec:lxrsub}). We find a range of $1.3\times10^{42}$ to $4.0\times10^{44}$~erg~s$^{-1}$ for our sample, which puts all objects firmly into the Seyfert regime. We note that one of our objects (NGC 7213) is classified as a broad-line LINER \citep{veroncetty2010}.

\begin{table*}
	\caption{Previously unpublished data \label{tab:newdata}}
	\begin{tabular}{@{}lccccccl}
		\hline
		\hline
		Object & Obs. ID & RA 		& Dec 					& Obs. Date 	& Slit Width & Slit Orientation \\
		 	   & 	     & [h m s] 	& [$^{\circ}$ $'$ $''$] & [YY-MM-DD] 	& [arcsec]	& [$^{\circ}$]\\
		\hline
		NGC 1068		& 075.B-0163(A)	& 02 42 40.7 &       -00 00 48 & 05-11-18	& 0.4	& 356	\\
		Circinus		& 280.B-5068(A)	& 14 13 09.9 &       -65 20 21 & 08-03-14	& 1.0	& 315	\\
		ESO138-G001		& 085.B-0251(A)	& 16 51 20.1 &       -59 14 05 & 10-06-27	& 0.8	& 320	\\
		F9				& 085.B-0251(B)	& 01 23 45.8 &       -58 48 21 & 10-07-20	& 0.8	& 45	\\	
		F49				& 085.B-0251(A)	& 18 36 58.3 &       -59 24 09 & 10-09-01	& 0.8	& 105	\\
		\hline
	\end{tabular}
\end{table*}

\begin{table*}
	\small
	\caption{Characteristics of our AGN mid-IR spectroscopic sample \label{tab:agnsample}}
	\begin{tabular}{@{}p{2.0cm}p{1.0cm}cccccccccccc}
		\hline
		\hline
		Object & Type\textsuperscript{a} & RA & Dec & $z$\textsuperscript{b} & \hphantom{1p} $D_L$\textsuperscript{c} \hphantom{1p} &
		Slit ori.\textsuperscript{d} & log \lx\textsuperscript{e}  &  \rsub\textsuperscript{f}  & Ref.\textsuperscript{g} & PAH  \\
		& & [h m s] & [$^{\circ}$ $'$ $''$] & & [Mpc] & [$^{\circ}$] &[erg/s] & [pc]  &  & detected \\
		\hline
           IZw1 &    S1h &   00 53 34.9 &   +12 41 36 &   0.0578 &   269.0 &     1 &   43.68\,$\pm$\,0.17 &    0.081$\,\pm\,0.021$ &   3 &   yes   \\
           F9 &   S1.2 &   01 23 45.8 &   -58 48 21 &   0.0466 &   215.0 &    45 &   43.99\,$\pm$\,0.11 &    0.111$\,\pm\,0.025$ &   1 &    no   \\
           NGC 1068 &    S1h &   02 42 40.7 &   -00 00 48 &   0.0030 &    14.4 &   356 &   43.64\,$\pm$\,0.30 &    0.078$\,\pm\,0.036$ &   1 &    no   \\
           NGC 1365 &   S1.8 &   03 33 36.4 &   -36 08 25 &   0.0051 &    17.9 &   303 &   42.12\,$\pm$\,0.20 &    0.017$\,\pm\,0.005$ &   3 &    no   \\
           LEDA 17155 &    S1h &   05 21 01.4 &   -25 21 45 &   0.0427 &   196.0 &   360 &   44.46\,$\pm$\,0.40 &    0.180$\,\pm\,0.130$ &   3 &   yes   \\
           NGC 2110 &    S1h &   05 52 11.4 &   -07 27 22 &   0.0080 &    35.9 &    55 &   42.67\,$\pm$\,0.10 &    0.029$\,\pm\,0.005$ &   2 &    no   \\
           ESO 428-G14 &     S2 &   07 16 31.2 &   -29 19 29 &   0.0063 &    28.2 &    60 &   42.15\,$\pm$\,0.37 &    0.017$\,\pm\,0.011$ &   2 &    no   \\
           MCG-5-23-16 &    S1i &   09 47 40.1 &   -30 56 55 &   0.0095 &    42.8 &    50 &   43.28\,$\pm$\,0.30 &    0.054$\,\pm\,0.025$ &   2 &    no   \\
           MARK 1239 &    S1n &   09 52 19.1 &   -01 36 43 &   0.0211 &    95.4 &   320 &   43.32\,$\pm$\,0.30 &    0.056$\,\pm\,0.026$ &   3 &    no   \\
           NGC 3227 &   S1.5 &   10 23 30.6 &   +19 51 54 &   0.0049 &    22.1 &   300 &   42.14\,$\pm$\,0.21 &      0.012$ \pm 0.001$ &   2 &   yes   \\
           NGC 3281 &     S2 &   10 31 52.1 &   -34 51 13 &   0.0118 &    52.8 &     5 &   43.27\,$\pm$\,0.37 &    0.054$\,\pm\,0.016$ &   3 &    no   \\
           NGC 3783 &   S1.5 &   11 39 01.7 &   -37 44 19 &   0.0108 &    48.4 &   315 &   43.24\,$\pm$\,0.07 &      0.071$ \pm 0.004$ &   2 &    no   \\
           NGC 4507 &    S1h &   12 35 36.6 &   -39 54 33 &   0.0128 &    57.5 &   330 &   43.21\,$\pm$\,0.17 &    0.050$\,\pm\,0.030$ &   2 &    no   \\
           ESO 323-G77 &   S1.2 &   13 06 26.1 &   -40 24 53 &   0.0159 &    71.8 &   354 &   42.76\,$\pm$\,0.12 &    0.032$\,\pm\,0.004$ &   2 &   yes   \\
           MCG-3-34-64 &    S1h &   13 22 24.4 &   -16 43 42 &   0.0176 &    79.3 &   315 &   43.33\,$\pm$\,0.48 &    0.057$\,\pm\,0.014$ &   2 &    no   \\
           MCG-6-30-15 &   S1.5 &   13 35 53.7 &   -34 17 44 &   0.0087 &    38.8 &    30 &   42.80\,$\pm$\,0.14 &    0.033$\,\pm\,0.006$ &   2 &   yes   \\
           IC 4329A &   S1.2 &   13 49 19.2 &   -30 18 34 &   0.0170 &    76.5 &   120 &   43.85\,$\pm$\,0.09 &      0.130$ \pm 0.080$ &   2 &    no   \\
           Circinus &    S1h &   14 13 09.9 &   -65 20 21 &   0.0019 &     4.2 &   315 &   42.26\,$\pm$\,0.28 &    0.019$\,\pm\,0.017$ &   1 &   yes   \\
           NGC 5643 &     S2 &   14 32 40.7 &   -44 10 28 &   0.0047 &    20.9 &    85 &   42.23\,$\pm$\,0.38 &    0.019$\,\pm\,0.004$ &   2 &   yes   \\
           NGC 5995 &   S1.9 &   15 48 24.9 &   -13 45 28 &   0.0257 &   117.0 &    60 &   43.45\,$\pm$\,0.15 &    0.064$\,\pm\,0.010$ &   2 &   yes   \\
           ESO 138-G001 &     S2 &   16 51 20.1 &   -59 14 05 &   0.0093 &    41.8 &   320 &   43.31\,$\pm$\,0.37 &    0.056$\,\pm\,0.023$ &   1 &   yes   \\
           F49 &    S1h &   18 36 58.3 &   -59 24 09 &   0.0199 &    90.1 &   105 &   43.34\,$\pm$\,0.11 &    0.057$\,\pm\,0.036$ &   1 &   yes   \\
           MARK 509 &   S1.5 &   20 44 09.7 &   -10 43 25 &   0.0335 &   153.0 &   315 &   44.12\,$\pm$\,0.09 &      0.125$ \pm 0.008$ &   2 &    no   \\
           IC 5063 &    S1h &   20 52 02.3 &   -57 04 08 &   0.0109 &    49.1 &    45 &   42.86\,$\pm$\,0.10 &    0.035$\,\pm\,0.008$ &   2 &    no   \\
           NGC 7213 &    S3b &   22 09 16.3 &   -47 10 00 &   0.0051 &    23.0 &   300 &   42.19\,$\pm$\,0.06 &    0.018$\,\pm\,0.011$ &   2 &    no   \\
           NGC 7469 &   S1.5 &   23 03 15.6 &   +08 52 26 &   0.0151 &    67.9 &   315 &   43.19\,$\pm$\,0.07 &      0.041$ \pm 0.001$ &   2 &   yes   \\
           NGC 7582 &    S1i &   23 18 23.5 &   -42 22 14 &   0.0044 &    23.0 &    60 &   42.53\,$\pm$\,0.38 &    0.025$\,\pm\,0.005$ &   2 &   yes   \\
           NGC 7674 &    S1h &   23 27 56.7 &   +08 46 45 &   0.0277 &   126.0 &   120 &   44.02\,$\pm$\,0.55 &    0.115$\,\pm\,0.025$ &   2 &   yes   \\
		\hline

	\end{tabular}
	\tablecomments{$^{(a)}$ AGN types from \citet{veroncetty2010}; $^{(b)}$ redshifts from NED; $^{(c)}$ luminosity distance based on CMB reference frame redshifts from NED and $H_0 = 73$ km s$^{-1}$ Mpc$^{-1}$, $\Omega_\mathrm{m} = 0.27$, and $\Omega_{\mathrm{vac}} = 0.73$; $^{(d)}$ slit orientation; $^{(e)}$ \lx = absorption-corrected 2-10\,keV luminosity from \citet{asmus2015}; $^{(f)}$ dust sublimation radius calculated from equation~\ref{eqn:lxrsublinmix} in this work, except for IC4329 A \citep{kishimoto2011}, NGC 3783 \citep{glass1992}, and MARK 509, NGC 3227, and NGC 4593 \citep{koshida2014}; $^{(g)}$ reference for the VISIR spectroscopic data.}
	\tablerefs{(1) This work; (2) \citet{honig2010}; (3) \citet{burtscher2013}.}
\end{table*}

\section{Extraction of the 11.3\mumeter PAH feature}\label{sec:method}
Ground-based mid-IR observations of PAH features provide the highest angular resolution. However, they are restricted to the atmospheric N-band window between 8 and 13\mumeter, which limits the choice of features that can be investigated to the 7.7/8.6\mumeter complex and the 11.3\mumeter PAH features. Of these two spectral regions, only the 11.3\mumeter feature is completely within the spectral range covered by VISIR. Therefore we spatially map the 11.3\mumeter PAH feature as a proxy for PAH molecules. We reiterate that this feature seems to be the one least suppressed by radiation around AGN \citep{smith2007,diamondstanic2010}. For each object, we extract spectra at various distances from the nucleus in steps of 0.4~arcsec and aperture widths of 0.4~arcsec, which corresponds to the typical spatial resolution of the data at this wavelength.

\begin{figure}
	\includegraphics[width=62mm,angle=90]{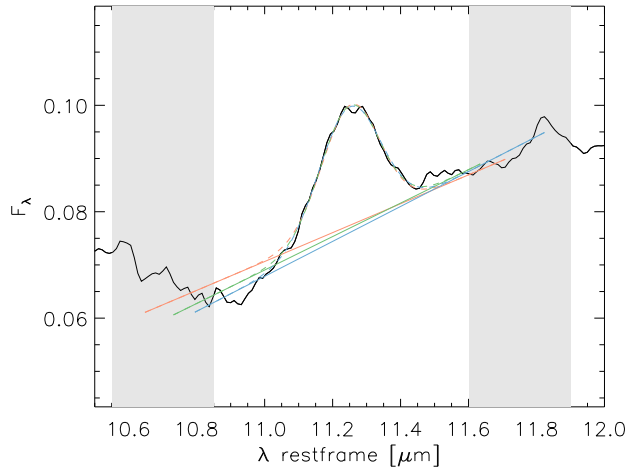}
	\caption{Three examples of fits to the 11.3\mumeter PAH feature in the spectrum extracted at an offset of 0.4~arcsec from the nucleus in NGC 3227. The shaded gray areas show the spectral regions from which the continuum anchors are drawn. Each fit is shown in its own color with the continuum as the solid line and the Gaussian model as the dashed line.}
	\label{fig:pahfits}
\end{figure}

The emitted PAH flux in each spectrum is calculated from a Gaussian model fit to the continuum-subtracted 11.3\mumeter emission. We note that other studies have used different functional forms for the PAH emission \citep[e.g.,][]{uchida2000,peeters2002,smith2007} but since we are only interested in the strength, not the shape, of the feature, a Gaussian function is sufficient for our purpose\footnote{It should be noted that the fitting procedure used makes the 11.3\mumeter feature strengths about half the ones derived with PAHFIT \citep[see Figure~8 and Table~6 in][]{smith2007}.}. The continuum is modelled by a local linear fit. This fitting routine is repeated 50 times for each spectral extraction by randomly varying the continuum anchors on each side of the PAH feature. Three examples of these Gaussian models are shown in Figure~\ref{fig:pahfits}. Such a Monte-Carlo-based procedure, similar to the one used by \citet{hernancaballero2011}, allows us to include the systematic uncertainty due to the selection of the continuum subtraction into the overall errors of the extracted lines. The continuum anchors are generally drawn from the regions between 10.55--10.85\mumeter and 11.6--11.9\mumeter in the rest frame. However, we varied these windows slightly if significant background contamination was present. For each repetition of the fitting routine we manually reject or accept the fit. The mean and standard deviation of all accepted fits for a given extraction is then used as the measured PAH flux and its uncertainty, respectively. In those cases where the signal-to-noise (S/N) ratio of the continuum drops below 1 or where no clear 11.3\mumeter PAH feature is present, we attempt to measure an upper limit. In a few cases, neither a measurement nor an upper limit can be extracted, simply because the spectrum is completely noise dominated or there is not enough flux in the spectrum.

As a note of caution, we risk underestimating the uncertainty of our fits by biasing the manually accepted fits to those that look similar. If the noise in our manually chosen continuum regions does not represent the true amount of continuum noise we also risk to underestimate the 11.3\mumeter PAH flux uncertainty. This is especially a concern if the continuum range is narrow, the spectral noise is low, or both. However, given the wide range of continuum shapes and noise levels, in addition to the narrow spectral range of our data, the outlined procedure is the most practical one.

\begin{figure*}
	\includegraphics[width=90mm,angle=270]{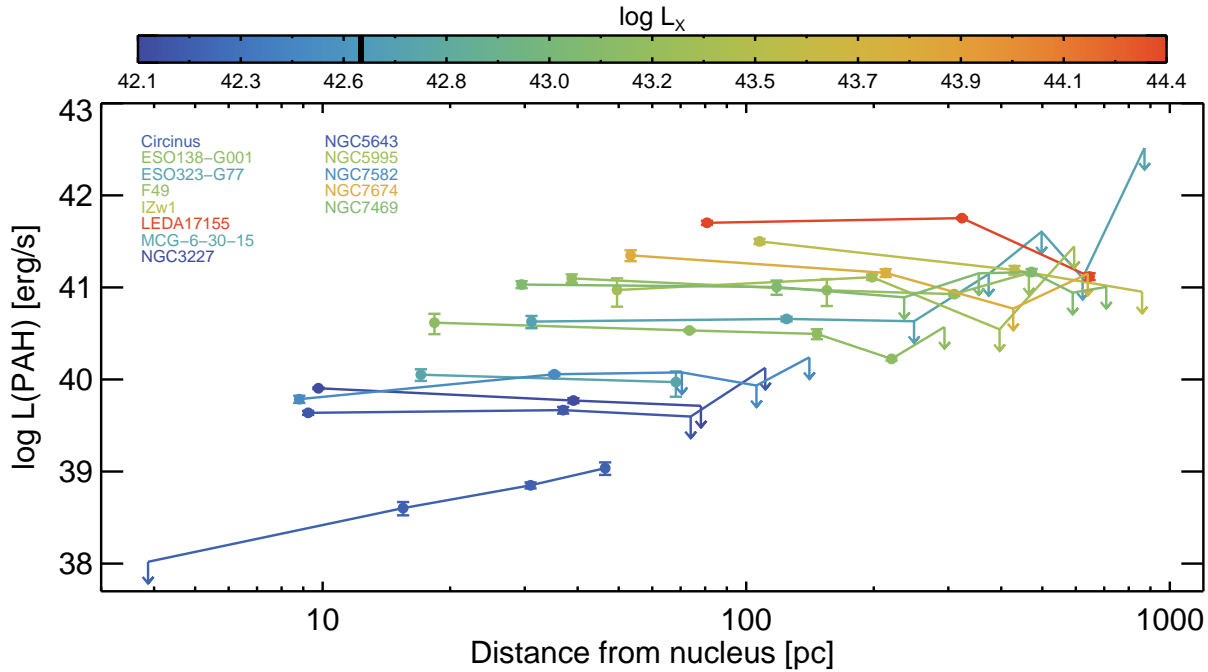}
	\caption{Luminosity of the 11.3\mumeter PAH feature measured within 0.4~arcsec apertures as a function of physical distance for the 13 AGN in our sample where we detect the feature. Upper limits are shown as downward arrows. Each object is colour coded according to its intrinsic X-ray luminosity \lx listed in Table\,\ref{tab:agnsample}.}
	\label{fig:pahvsparsec}
\end{figure*}

We find PAH features in 13 out of the 28 objects. The PAH luminosity as a function of physical distance from the AGN for these 13 objects is shown in Figure~\ref{fig:pahvsparsec}. This detection rate is similar to what is found in other studies of Seyfert galaxies \citep[e.g.,][]{esquej2014,alonsoherrero2016}. For Circinus and Mark 1239, we do not detect the 11.3\mumeter PAH feature in the central aperture, but the feature appears at apertures further away from the nucleus. For Circinus, this is consistent with the work by \citet{roche2006} who found the 11.3\mumeter PAH feature to be very weak within 2~arcsec of the nucleus.  

Figure~\ref{fig:pahnondetec} shows upper limits for the 15 AGN where we do not detect the 11.3\mumeter PAH feature. Most of these are within the region (gray shaded area) where we detect PAH features for the remaining 13 objects in our sample, which means that the PAH features might very well be present in these objects, but that they are too weak for us to detect them. For the few remaining sources, we can only speculate as to why we do not find any 11.3\mumeter PAH feature. One possibility is that the molecular clouds are not sufficiently self-shielding to protect the PAH molecules from destruction by the hard AGN radiation field. This could be a matter of either not dense enough clouds or too few clouds, meaning a too low covering factor, or both. As an example, work by \citet{mason2009} shows that modeling the silicate emission of NGC 2110 favors models with only a few clouds along the line of sight, probably not enough to provide sufficient self-shielding for the PAH grains to survive in the vicinity of the AGN. We stress that these suggestions only are speculative and need further investigation in the future.

For fifteen of the objects in our sample, the 11.3\mumeter PAH feature flux and equivalent width have been measured previously by \citet{esquej2014} using spectra extracted with aperture sizes between 0.35 and 1.00~arcsec (except for Circinus where they used a 3.7~arcsec aperture). For ESO323-G77, we detect the 11.3\mumeter PAH feature whereas \citet{esquej2014} only provide an upper limit. Our measurement is consistent with this upper limit. For two objects, NGC 1068 and Mark 509, \citet{esquej2014} detect the 11.3\mumeter PAH feature, whereas we are unable to reliably measure a flux. For these two objects, we find that the contrast between the feature and the underlying continuum in the central aperture is too low for a detection within our extraction process. For the rest of the objects common to both studies, detections and non-detections agree. Figure~\ref{fig:pahspectra1}, Figure~\ref{fig:pahspectra2}, and Figure~\ref{fig:pahspectra3} in Appendix~\ref{app:notes} show the spectra extracted inside the central 0.4~arcsec aperture for all the objects in our sample. 

In Appendix~\ref{app:psfcheck}, we show the emitted surface flux of the 11.3\mumeter PAH feature and the underlying continuum for the 13 objects where we detect the feature. For all these objects, the emitted continuum surface flux -- originating from the unresolved nuclear source -- declines more rapidly than that of the 11.3\mumeter PAH feature. A similar behaviour was found by \citet{alonsoherrero2014} for a few local AGN observed with the Gran Telescopio CANARIAS CanariCam. This means that the PAH features are indeed spatially resolved, at least marginally. However, this also implies that the point spread function (PSF) will affect the radial profiles of the PAH features and needs to be taken into account when modelling/interpreting the results. All the extracted spectra for the 28 objects are made available online\footnote{Will be made available on VizieR upon publication}.

\section{The \lxrsub correlation}\label{sec:lxrsub}
In our data analysis and discussion, we want to assess if the AGN has any influence on the PAH feature in its vicinity by searching for evidence that the PAH strength in the nucleus is (universally) related to that of the AGN. However, our AGN sample spans a large range in luminosity and distance: the same angular scale corresponds to a different degree of incident AGN flux, since both the physical distance from the AGN and the AGN intrinsic luminosity determine its strength in each of our extraction windows. The difference in physical scales (=pc; as shown in Fig.~\ref{fig:pahvsparsec}) can easily be accounted for by taking into account the physical distances from the AGN for each of our extraction windows in each source. For the differences in incident AGN radiation, we have to further normalise the extractions windows according to a scale that accounts for the $1/r^2$ dilution of radiation. We thus normalise the emission profile by a scale intrinsic to the AGN allowing us to directly compare the objects.

\begin{figure}
	\includegraphics[width=45mm,angle=270]{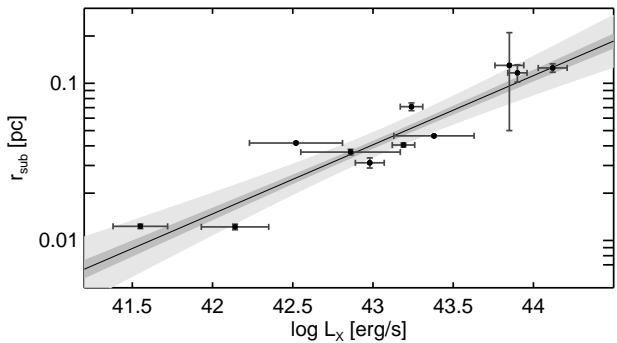}
	\caption{Dust sublimation radius (\rsub) as a function of intrinsic 2--10~keV X-ray luminosity (\lx) for the 11 type 1 AGN summarised in Table~\ref{tab:lxrsub}. Solid line shows the best power-law fit using \texttt{linmix\_err}. Shaded areas shows the 68 and 95~per~cent confidence intervals, respectively.}
	\label{fig:lxvsrsub}
\end{figure}

Since the PAH features are supposedly originating from dense molecular and dusty clouds, a convenient choice for such an intrinsic scale is the dust sublimation radius. Unfortunately, only few of the type 1 objects in our sample have measured sublimation radii from near-IR interferometry or reverberation mapping \citep{glass1992,kishimoto2009,kishimoto2011,koshida2014}. Moreover, the current empirical relations between AGN optical luminosity and hot dust time lags are only applicable to type 1 AGN, since type 2s suffer from significant obscuration in the optical. Therefore we take a different approach and aim at establishing a relation between an intrinsic AGN luminosity tracer that is unaffected by obscuration and the near-IR time lags. For that we compile a sample of 11 sources that have both near-IR time lag measurements and absorption-corrected 2--10~keV X-ray luminosities (\lx). These are listed in Table~\ref{tab:lxrsub} and the data are visualised in Fig.~\ref{fig:lxvsrsub}. Using X-rays has the advantage that intrinsic X-ray luminosities can be estimated even for highly obscured sources up to the Compton-thick limit, and beyond, if one can robustly model the broadband X-ray spectra \citep[for a discussion of uncertainties in intrinsic X-ray luminosities of Compton-thick AGN, see for example][]{gandhi2014,gandhi2015}, which allows us to reliably estimate the sublimation radius in optically-obscured type 2 AGN. 

\begin{table*}
	\caption{Objects used to calibrate the \lxrsub relation \label{tab:lxrsub}}
	\begin{tabular}{@{}lccccccl}
		\hline
		\hline
		Object & \rsub & $\sigma$(\rsub) & Ref. & log $L_X$\textsuperscript{a} & $\sigma$(log $L_X$)\textsuperscript{b} & Ref. \\
		& [pc]  & [pc] & & [erg/s] & [erg/s] & \\
		\hline
		Akn 120								& 0.1165	& 0.0147	& 1	& 43.90		& 0.06		& 4,\,5,\,6	\\	
		IC 4329A\textsuperscript{c}			& 0.13		& 0.08		& 2	& 43.85		& 0.09		& 7,\,8,\,9,\,6 \\
		MARK 509							& 0.125		& 0.008		& 1	& 44.12		& 0.09		& 10,\,11,\,12	\\
		MARK 590							& 0.0312	& 0.0023	& 1	& 42.98		& 0.09		& 13,\,14,\,15	\\	
		NGC 3227							& 0.0122	& 0.0005	& 1	& 42.14		& 0.21		& 16,\,8,\,12,\,6	\\
		NGC 3783							& 0.071		& 0.004		& 3	& 43.24		& 0.07		& 17,\,18,\,19,\,6	\\
		NGC 4051							& 0.0123	& 0.0004	& 1	& 41.55		& 0.17		& 20,\,21,\,22,\,6	\\
		NGC 4151							& 0.0417	& 0.0006	& 1	& 42.52		& 0.29		& 18,\,23,\,24	\\	
		NGC 4593							& 0.0365	& 0.0015	& 1	& 42.86		& 0.31		& 25,\,18,\,26,\,6	\\
		NGC 5548							& 0.0463	& 0.0006	& 1	& 43.38		& 0.25		& 27,\,28,\,18	\\	
		NGC 7469							& 0.0405	& 0.0012	& 1	& 43.19		& 0.07		& 29,\,17,\,18	\\
		\hline
	\end{tabular}
	\tablecomments{$^{(a)}$ \lx = absorption-corrected 2-10\,keV luminosity from \citet{asmus2015} with the original references from that work listed in the last column of this table; $^{(b)}$ $\sigma$(log $L_X$) = uncertainty on \lx adopted from \citet{asmus2015}; $^{(c)}$ we assign an uncertainty of 0.2 dex to \rsub for IC4329A based on the scatter in Figure 4 of \citet{kishimoto2007} between radial size and UV-luminosity and Figure~30 in \citet{suganuma2006} between time lag (i.e., radial size) and $V$-band magnitude as quoted by \citet{kishimoto2011}.} 
	\tablerefs{(1) \citet{koshida2014}; (2) \citet{kishimoto2011}; (3) \citet{glass1992}; (4) \citet{vasudevan2009}; (5) \citet{winter2012}; (6) \citet{winter2009a}; (7) \citet{dadina2007}; (8) \citet{brightman2011a}; (9) \citet{brenneman2014}; (10) \citet{shinozaki2006}; (11) \citet{ponti2009}; (12) \citet{shu2010a}; (13) \citet{gallo2006}; (14) \citet{longinotti2007}; (15) \citet{rivers2012}; (16) \citet{markowitz2009}; (17) \citet{nandra2007}; (18) \citet{rivers2011a}; (19) \citet{brenneman2011}; (20) \citet{king2011}; (21) \citet{mchardy2004}; (22) \citet{vaughan2011}; (23) \citet{wang2010}; (24) \citet{lubinski2010}; (25) \citet{horst2008}; (26) \citet{markowitzreeves2009}; (27) \citet{andradevelazquez2010}; (28) \citet{krongold2010}; (29) \citet{asmus2015}.}
\end{table*}

We model the data with a power-law of the form $r_{sub} \propto L^\gamma$ using the IDL procedure \texttt{linmix\_err} \citep{kelly2007}. The underlying method uses a Bayesian approach to linear regression and accounts for effects of uneven sampling. The data and best-fit linear regression (solid black line) are shown in Figure~\ref{fig:lxvsrsub}. The best-fit model parameters are
\begin{equation}
\label{eqn:lxrsublinmix}
r_{\mathrm{sub}} = (0.0407\,\pm\,0.0003)\, \mathrm{pc} \left(\frac{L_\mathrm{X}}{10^{43}\,\mathrm{erg/s}}\right)^{0.44\,\left(^{+0.06}_{-0.07}\right)}
\end{equation}
with a correlation strength measured in terms of the Spearman rank of $\rho_s = 0.97\,^{+0.02}_{-0.06}$. The power-law slope of $0.44^{+0.06}_{-0.07}$ is consistent with the canonical $1/2$ within the 68~per~cent confidence interval \citep{barvainis1987,suganuma2006,kishimoto2007}. Since the relation between X-ray and UV-optical luminosity is observed as non-linear \citep[e.g.,][]{lusso2010,marchese2012}, a deviation from $1/2$ may however be expected. The bisector regression of the $L_\mathrm{disc}-L_\mathrm{2-10\,keV}$ relation of \citet{marchese2012} implies a $r_\mathrm{sub}-L_\mathrm{x}$ power-law index of $\sim0.42\pm0.03$, which is consistent with our measurement.

The luminosity range spanned by the 11 objects in Table~\ref{tab:lxrsub} is almost identical to that of the 28 objects of our main sample. Together with the fact that we use the intrinsic X-ray luminosity, this justifies our use of equation~\ref{eqn:lxrsublinmix} to estimate dust sublimation radii for both type 1 and type 2 AGN, at least for objects that are not Compton-thick. We note that some of our sources are Compton-thick but that we consider their \lx measurements to be very reliable as they are shown to lie on the same \lxlmir correlation as those sources that are not Compton-thick \citep{gandhi2009,asmus2015}. In addition, detailed broadband X-ray spectroscopy of several of these has found excellent agreement with expectations based upon the correlation (cf. \citet{bauer2015} for NGC 1068; \citet{arevalo2014} for Circinus; \citet{annuar2015} for NGC 5643).

\section{Results and analysis}\label{sec:results}

\subsection{Qualitative analysis}\label{sec:results_qual}

In Fig.~\ref{fig:pahvsparsec}, we show the radial dependence of the PAH luminosities, scaled in physical units. As can be seen, the absolute scaling is offset by more than 3 orders of magnitudes. The physical scales of detected fluxes range from 8\,pc to 700\,pc (i.e. 2 orders of magnitudes), which illustrates the increased spatial resolution of about a factor of 10 as compared to Spitzer IRS observations. Given the large offset in luminosities, it is difficult to make a comparison between the different objects from this plot. In particular, if we are interested in the possible effect of the AGN on the PAHs, we want to make sure that the physical scales and the observed fluxes of all the objects are normalised for the difference in AGN luminosities. Therefore, we normalise the physical scales by an AGN-intrinsic scale -- here the dust sublimation radius as defined in section~\ref{sec:lxrsub}. We also use the same intrinsic scale to normalise the observed PAH fluxes or luminosities to the emitted surface flux from the respective spatial scale. This way, any given distance from the AGN receives the same radiation field from the AGN in all sources and the emitted radiation can directly be compared among the different objects.

\begin{figure*}
	\includegraphics[width=90mm,angle=270]{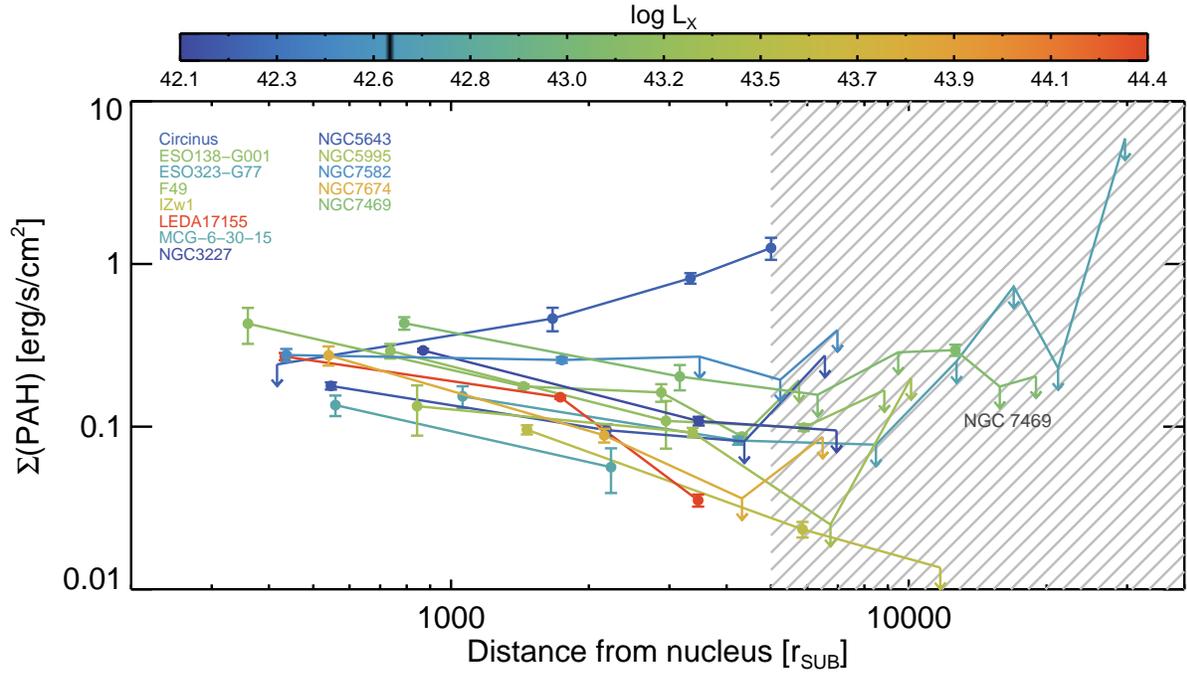}
	\caption{Emitted surface flux of the 11.3\mumeter PAH feature measured within 0.4~arcsec apertures as a function of sublimation radius for the 13 AGN in our sample where we detect such features. Upper limits are shown as downward arrows. The hashed gray area above 5000\,\rsub shows where emission associated with star formation starts to become the dominant excitation mechanism for the PAH grains (see \S~\ref{sec:results}). Each object is colour coded according to its intrinsic X-ray luminosity \lx.}
	\label{fig:pahvsrsub}
\end{figure*}

\begin{figure*}
	\includegraphics[width=90mm,angle=270]{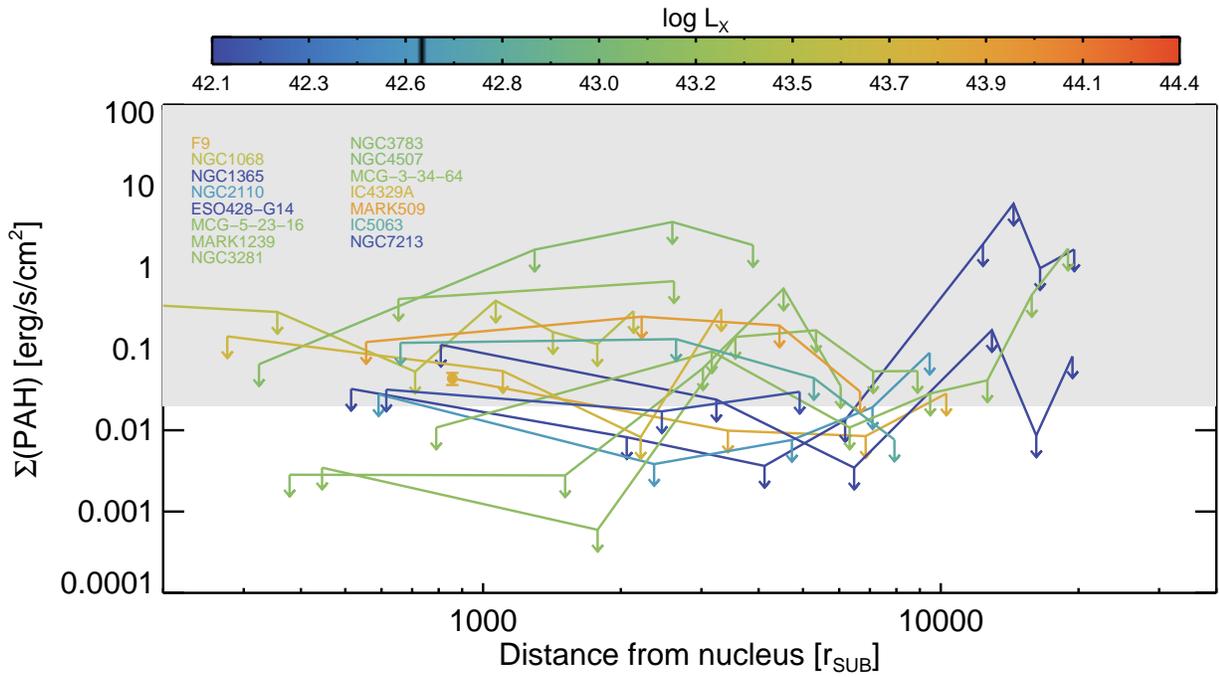}
	\caption{Upper limits for the emitted surface flux of the 11.3\mumeter PAH feature measured within 0.4~arcsec apertures as a function of sublimation radius (see \S\,\ref{sec:lxrsub}) for the 15 AGN in our sample where we do not detect the feature. The shaded gray area shows where our detections fall (see Figure~\ref{fig:pahvsrsub}) and indicate the detection limit for the 11.3\mumeter PAH feature. Each object is colour coded according to its intrinsic X-ray luminosity \lx.}
	\label{fig:pahnondetec}
\end{figure*}

Figure~\ref{fig:pahvsrsub} shows this renormalised, emitted 11.3\mumeter PAH surface flux for the 13 objects where we detect this feature. These were calculated from the extracted PAH fluxes and the respective aperture area. The first interesting observation of this renormalisation is that the overall range in observed PAH fluxes is reduced from more than 3 orders of magnitudes in the physical scaling to less than 1 order of magnitude in the AGN-normalised scaling. If the PAH emission would be independent of the AGN, we would not necessarily expect that scaling for the AGN would reduce the scatter in the observed flux normalisations of the sample, unless the star formation closely correlates with the AGN activity in all galaxies. Therefore, the source of PAH excitation must be related to the AGN power, i.e., it could either be a physical process that scales with the AGN power or it could be the AGN itself. We will discuss these scenarios further in section~\ref{sec:cloudy}.

The second interesting observation of this AGN-normalised PAH emission concerns the dependence on distance. In Fig.~\ref{fig:pahvsrsub}, we see that all objects except for Circinus fall into a narrow band where the PAH surface flux decreases with increasing distance from the AGN. This confirms the earlier finding by \citet{alonsoherrero2014} for Seyferts and LIRGs. The band only widens at distances larger than $\sim$5,000\,\rsub (marked by the grey-hatched area), although part of this widening is related to non-detections rather than detections. A prominent example of the change in behaviour at distances $>$5,000\,\rsub is NGC 7469 (green labelled curve in Figure\,\ref{fig:pahvsrsub}). This source has a well known starburst ring at about 2~arcsec distance from the AGN \citep{genzel1995,diazsantos2007}, which coincides with this turnover. As such, the turnover in this source indicates that local star formation activity clearly is the dominant contributor to the PAH emission at these radii. We will statistically analyse the scatter in the slopes of all the objects in the next subsection.

Circinus is an outlier both in terms of radial slope as well as absolute PAH surface flux. \citet{tristram2014} have shown that it has a strong optical depth gradient of $\Delta\tau = 27$~arcsec$^{-1}$ in the mid-IR from the west to the east side of the nucleus on parsec scales. Such a gradient was also found by \citet{roche2006}, albeit with a value of only $\Delta\tau = 0.6$~arcsec$^{-1}$ and on larger scales. Larger scale observations indicate that the eastern side of the galactic disc \citep{freeman1977} and a circumnuclear molecular ring \citep{curran1998,curran1999} are closer to us, so that the obscuration increases towards the nucleus \citep[see also][]{mezcua2016}. This provides a feasible explanation for the reverse slope we observe for the PAH features in Circinus compared with the rest of our sample. Indeed, we find a slope of the PAH features in Circinus which corresponds to a gradient in the optical depth of $\Delta\tau = 2.5$~arcsec$^{-1}$ at mid-IR wavelengths, in between the two previous measurements. Therefore, we consider Circinus as a special case and leave it out of the following analysis. This leaves us with 12 objects to investigate quantitatively.

\subsection{Statistical analysis}\label{sec:statan}

We want to quantify the slope of the radial dependence of the PAH emission by a model of the form $\log \Sigma_{PAH} = \log \Sigma_{0,\mathrm{PAH}} + \alpha * \log(r_{\mathrm{sub}})$. For that, we first model the emitted PAH fluxes for each object individually and then combine the results to obtain a joint probability distribution function for the power-law slope. Since some objects show changes in the slope at larger radii, we also investigate how selection of an outer cut-off affects the fit. We perform a linear regression in log-log space using radial cut-offs at 3000, 5000, 6000, 7000, and 10,000~\rsub, respectively (see Table~\ref{tab:pahslope}). When modelling the PAH emission for each object, we generate 10,000 random realisations of the data using Monte Carlo simulations to account for the measurement uncertainties. We assume all measurements to have normally distributed errors and include the upper limits in the fit by allowing them to take on any value between the measured upper limit and 1~per~cent of this value.

Because the point spread function of the observations smears out the signal and distributes flux from smaller to larger radii, we observe a shallower PAH slope than the intrinsic one. To account for this in our fitting process, we convolve the power-law model with a Gaussian function with FWHM = 0.4~arcsec, corresponding to the spatial resolution of the observations. Therefore, the fitted slopes for each object represent the intrinsic dependence, which we will refer to in the following, unless otherwise stated. 

We combine the results of the modelling process for each object to get a joint probability distribution function for the power-law slope $\alpha$. From this distribution we extract the median slope, its 68~per~cent confidence interval, and the sample scatter of the fitted slopes. As the resulting probability distribution for the slope depends slightly on how we bin the discrete distributions for $\alpha$, we calculate results for bin sizes of 0.01 and 0.1, respectively. These joint probability distribution functions are shown in Appendix Fig.~\ref{app:jpds}. As we require at least three data points to be included within the radial cut-off for each object, the number of objects included in the estimation of the PAH slope increases with radial cut-off. The power-law slopes for each object and each radial cut-off are shown in Table~\ref{tab:pahslopeindividual} in Appendix~\ref{app:jpds}.

\begin{table}
	\caption{Slope of the \rsub vs. \pahsurfflux relation\label{tab:pahslope}}
	\begin{tabular}{@{}cccccc}
		\hline
		\hline
		$r_{sub,outer}$ & $\alpha_{bin=0.01}$\textsuperscript{a} & $\alpha_{bin=0.1}$\textsuperscript{b} & $\sigma_\alpha$\textsuperscript{c} & $n_{objects}$ \\
		\hline
		 3000 & $-0.69^{+0.33}_{-0.20}$ & $-0.8^{+0.3}_{-0.3}$ & 0.32 & 2 \\
		 5000 & $-1.19^{+0.10}_{-0.05}$ & $-1.2^{+0.2}_{-0.1}$ & 0.42 & 5 \\
		 6000 & $-1.16^{+0.06}_{-0.04}$ & $-1.1^{+0.1}_{-0.1}$ & 0.36 & 7 \\
		 7000 & $-1.15^{+0.04}_{-0.04}$ & $-1.1^{+0.1}_{-0.1}$ & 0.40 & 9 \\
		 10000 & $-1.14^{+0.04}_{-0.03}$ & $-1.0^{+0.1}_{-0.1}$ & 0.35 & 10 \\
		\hline
	\end{tabular}	
	\tablecomments{$^{a}$ Median slope and $\pm 1 \sigma$ uncertainty using a bin size \\of 0.01 for the joint probability distribution; $^{b}$ median slope and \\$\pm 1 \sigma$ uncertainty using a binsize of 0.1 for the joint probability \\distribution; $^{c}$ scatter of the fitted slopes. For a radial cutoff \\between 5000 and 7000 \rsub, the spearman rank significance for \\most objects is $<-0.8$.}
\end{table}

Table~\ref{tab:pahslope} shows our results for the joint PAH slope using the different radial cut-offs. For radial cut-offs between 5,000 and 10,000~\rsub, the slope is similar with a value of approximately $-1.1\pm0.1$, independent of the bin size used. For the radial cut-off of 3,000 \rsub the slope is very shallow and not consistent with the other cut-offs. However, this cut-off region includes only two objects, which means that the slope is quite uncertain. Indeed, the slope is still consistent within 1.4$\sigma$ with all other estimates.

\begin{figure}
	\includegraphics[width=45mm,angle=270]{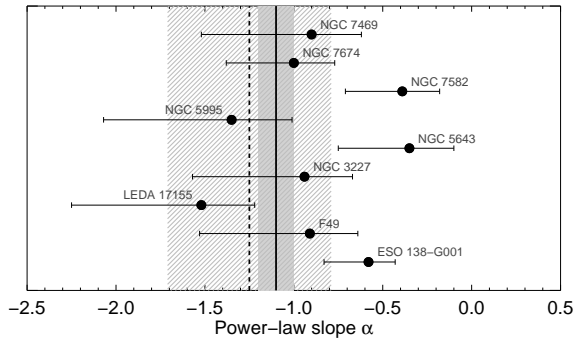}
	\caption{Comparison of individual PAH power-law slopes (filled black circles) and the joint PAH slope (solid black line) and its uncertainty (filled gray area) for a radial cut-off of 7000 \rsub. Also shown is the power-law slope fitted to the CLOUDY models of the PAH emission (dashed black line) and its uncertainty (hatched gray area) described in \S~\ref{sec:cloudy_model}.}
	\label{fig:pahslopes}
\end{figure}

Figure~\ref{fig:pahslopes} shows the individual power-law slopes for the 9 objects within the radial cut-off at 7,000 sublimation radii (solid black circles). We also plot the 68\% confidence region of each fitted slope. The joint PAH slope of all 9 objects is shown as a solid black line with the 68\% confidence region as a solid gray background. 

\begin{table}
	\caption{Reduced $\chi^2$ statistic for consistency with the single joint slope (second column) and the slope extracted from the CLOUDY models (fourth column). Also listed is the intrinsic scatter in the slopes that needs to be added to receive a reduced $\chi_r^2 = 1$ for the joint slope (third column).}\label{tab:chi2}
	\begin{tabular}{@{}cccc}
		\hline
		\hline
		$r_{sub,outer}$ & $\chi_r^2$ joint slope & $\sigma_\mathrm{int}(\chi_r^2=1)$ & $\chi_r^2$ model \\ 
		\hline
3000 &  0.28 & $\ldots$ & 1.31 \\
5000 &  1.08  & 0.13 &   0.62\\ 
6000 &  1.74      &      0.19 &        0.87\\
7000 &  1.76      &      0.21  &       0.90\\
10000 &  1.66   &          0.20 &         1.25\\
\hline
	\end{tabular}	
	\tablecomments{The number of degrees of freedom are the \\ number of objects as listed in Table~\ref{tab:pahslope} minus 1.}
\end{table}

To judge if a joint slope is a reasonable hypothesis, we test consistency of the individual radial slopes with the nominal joint slope of $-1.1$. For the different outer cut-offs listed in Table~\ref{tab:pahslope}, we calculate the reduced $\chi_r^2$ statistic taking into account the asymmetric errors of the individual slopes and uncertainty in the common slope. The results are shown in Table~\ref{tab:chi2}. These values may be roughly interpreted as being consistent with \textit{no} joint slope at the $0.5-1.7\sigma$ level assuming normally distributed errors. We also test what intrinsic scatter of the population would be required to obtain $\chi_r^2=1$. These values are listed in the third column of Table~\ref{tab:chi2}. For comparison, the errors of the individual slopes are given in Appendix Table~\ref{tab:pahslopeindividual} and are larger than the required intrinsic scatter. We address expectations for the intrinsic scatter in Sect.~\ref{sec:cloudy_model}.

\section{Discussion: PAH excitation from a compact central radiation source}\label{sec:cloudy}

The similarity in the relation between PAH surface flux and luminosity-normalised distance from the AGN is not straight-forward to interpret. In a generic scenario where PAH emission only traces star formation, and where star formation and AGN activity are not related, all galaxies should have their own distribution of PAH-emitting gas, depending on the individual star-formation history, without characteristics in common across the sample. Here, we do find a common slope, which would make this scenario unlikely and requires some common physical property to cause the observed radial distribution of the PAH emission.

Since PAH emission is commonly believed to be associated with star formation, one possibility is a common distribution of star formation around the nucleus. This would require a common mechanism to trigger a universal distance-dependent star formation distribution in the black hole environment, which may be difficult to achieve. On the other hand, we find that the scatter in absolute values of the PAH fluxes reduces overall if we scale with AGN luminosity. Therefore, both the distribution of star formation as well as its strength need tight coupling to the luminosity of the AGN.

An alternative to circumvent the required spatial and energetic relations between AGN and stars is to decouple PAH emission from star formation. This way, it is not necessary to find a physical mechanism that causes a universal radial star formation distribution, but rather a common mechanism for the excitation of PAH emission. Here, we propose that this excitation source should be centrally located and compact with respect to the PAH emission. Under this hypothesis, the main driver for the common radial slope is the $r^{-2}$-dilution of radiation from the central source of radiation. In principle, this could still lead to different radial slopes: the PAH emission is effectively reprocessed radiation that traces the combination of (diluted) incident radiation and radial distribution of PAH-emitting material. Therefore, the PAH-emitting gas should be fairly uniformly distributed in the central environment, e.g. gas clouds in the ISM. The range in absolute surface flux values, i.e. the relative offsets in Fig.~\ref{fig:pahvsrsub}, is then caused by the volume filling factor of the PAH emitting gas clouds, which can be quite different from object to object. This hypothesis would explain both why we see PAH detections and non-detections among our sample and why those with detections show a common radial slope.

The key questions for this hypothesis are: Is it possible to reproduce the observed slopes? And, does any central radiation source provide enough energy to explain the observed emission? We consider the AGN and a central stellar cluster as the two major viable candidates for the central excitation source. In the following, we will first investigate the AGN by building a CLOUDY model for the proposed scenario and test the slopes and emitted fluxes under the influence of radiation from the AGN. Afterwards, we will discuss the energetic viability of a central stellar cluster as the source of excitation based on observational constraints from two of the objects in our sample.

\begin{figure*}
	\includegraphics[width=140mm,angle=0]{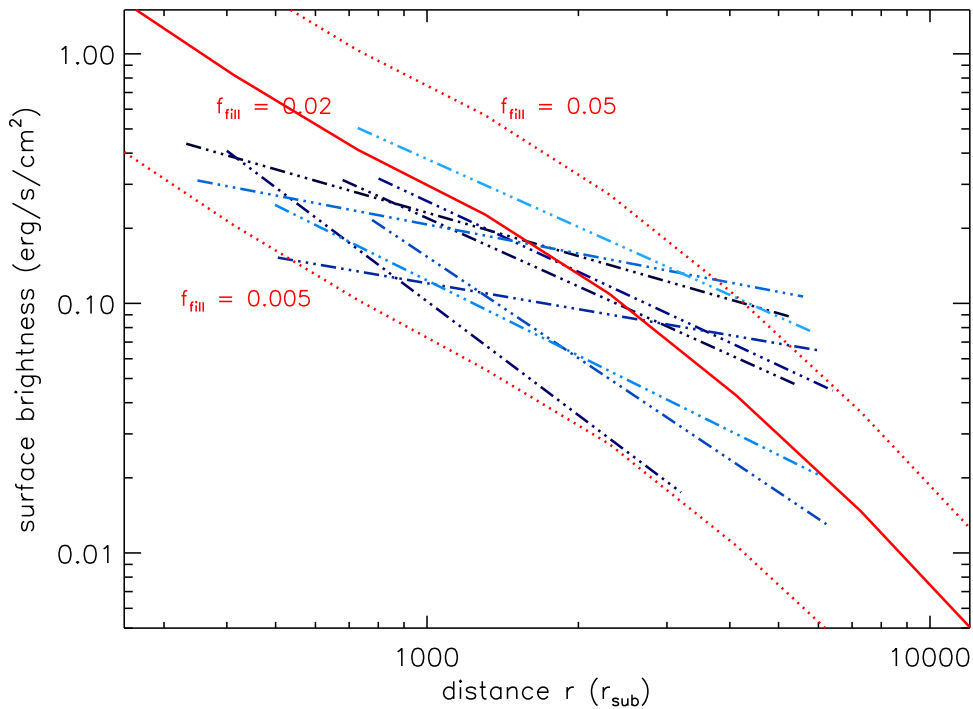}
	\caption{Comparison between power-law models of the observed emitted PAH fluxes (dot-dashed lines) and the CLOUDY models (red solid and dotted lines). The slopes of the power-law models and the CLOUDY models appear very similar.}
	\label{fig:pahvsrsubcloudy}
\end{figure*}

\subsection{AGN as the excitation source}\label{sec:cloudy_model}

We set up a CLOUDY\footnote{Calculations were performed with version C13.1 (March 2013) of CLOUDY, last described by \citet{ferland2013}.} model to test the hypothesis that the AGN acts as a central excitation source for uniformly distributed PAH-emitting gas in the central region. First, we calculate the emission of clouds at a range of distances between approximately 100 and 10\,000 \rsub from the AGN. The CLOUDY simulations take into account gas and dust physics and were set up to include back-heating effects on each cloud by surrounding clouds. This provides us with distance-dependent source functions $S_\nu(r)$ of the dusty gas clouds. We then combine these source functions to receive the distance-dependent emission from the central region as
\begin{equation}
F_\nu(r) \ dr = f_\mathrm{fill} \ S_\nu(r) \ r \ dr / (\pi r_\mathrm{sub}^2),
\end{equation}
where $f_\mathrm{fill}$ is the filling factor, which is the only free parameter in this model. For a comparison to observations, we extract the $11.3\,\micron$ PAH flux similar to the observations, i.e. by subtracting the linearly interpolated continuum underneath the observed feature (interpolation anchors are set at 11.11 and 11.63\mumeter, where the CLOUDY SEDs are continuum only).

In Fig.~\ref{fig:pahvsrsubcloudy}, we show the intrinsic PAH emission slopes of the 9 objects included for the cut-off at 7,000\,\rsub (see section~\ref{sec:statan}). To avoid crowding, we only plot the best-fit value of the slopes; the 68\% confidence regions are shown in Fig.~\ref{fig:pahslopes}. Overplotted in Fig.~\ref{fig:pahvsrsubcloudy} are model PAH surface fluxes with a range of model $f_\mathrm{fill}$ values that bracket these intrinsic slopes. Two lessons can be learned from this comparison: First, the required filling factor $f_\mathrm{fill}$ in the range of spatial scales covered by the observations is much lower than unity, which implies that the AGN does indeed provide enough energy to excite the observed PAH emission at the observed distances. Indeed, with $f_\mathrm{fill} \sim 0.01$, only a small number of AGN-heated, PAH-bearing dusty gas clouds are required to account for the observed fluxes. Alternatively, one may interpret the $f_\mathrm{fill}$ values in a way that only a small fraction of AGN light needs to reach out to these distances to provide enough energy for producing the observed PAH emission.

Second, we see that the model does not show one unique slope but becomes steeper at large distances from the AGN. To compare the model to observations, we extract model PAH fluxes at similar apertures as the observations and linearly fit slopes to these model-extracted fluxes. To account for the different intrinsic spatial scales in the objects, we vary the inner radius using a Monte Carlo scheme, so that the total spatial range from 400\,\rsub to 7,000\,\rsub is covered. We recover a model slope of $\alpha_\mathrm{mod} = -1.25 \pm0.46$. This slope is fully consistent with the joint slope of $\alpha=-1.1\pm0.1$. Similarly to the analysis in Sect.~\ref{sec:statan}, we also calculate $\chi_r^2$ values to quantify consistency of the observed slopes with the model. Here, we take into account the errors of the observations \textit{and} the intrinsic scatter of the model. The resulting $\chi_r^2$ statistic is listed in the fourth column of Table~\ref{tab:chi2}. We find $\chi_r^2$ that support the CLOUDY model as a reasonable representation of the data. Indeed, $\chi_r^2<1$ for the range of outer cut-offs between 3000 and 7000\,$r_\mathrm{sub}$. This is easily understood when considering that the intrinsic scatter of the model slope of 0.46 is larger than the one implied by testing each individual source against the joint slope.

In the models shown in Fig.~\ref{fig:pahvsrsubcloudy}, we adopt a Hydrogen column density of $N_H = 10^{23}\,\mathrm{cm^{-2}}$. This column density provides enough self-shielding within the dusty gas clouds for PAHs to survive. Such column densities can be expected from physical arguments of the state of clumps in the region of the torus \citep{krolikbegelman1988,vollmerbeckert2002,beckertduschl2004,honigbeckert2007,stern2014a,namekata2014}. Increasing the column density changes the model emission only marginally.

We would like to point out again that the AGN-excited PAH model we use contains only one free parameter, yet it reproduces the observed slope and is energetically viable. One may consider a slight variation of this model to allow for a slight variation in slopes by making the covering factor distance-dependent. However, this is not the purpose here. We only want to test the general viability of AGN as the heating source of the small-scale circumnuclear environment.

\subsection{Excitation by nuclear star clusters}\label{sec:stars}

A second source of central heating/excitation of the PAH molecules might be radiation from a compact nuclear star cluster (NSC). Here, compact would mean that the bulk of the radiation is originating from scales smaller than the observed and resolved PAH emission. Since NSCs may vary significantly among different galaxies, we assess their viability to produce the observed PAH emission via energy conservation arguments in two galaxies within our sample where observational constraints are available.

\noindent\textit{NGC 3227.} \citet{davies2006} report a total luminosity of the NSC in NGC 3227 of $L_\mathrm{NSC}(\mathrm{NGC~3227}) \sim 1 \times 10^{43}$~erg~s$^{-1}$ and measured in a $0.8$\,arcsec aperture. This corresponds to an intrinsic aperture radius of $\sim3300\,r_\mathrm{sub}$. In addition, they estimate the AGN luminosity as $L_\mathrm{AGN}(\mathrm{NGC~3227}) \sim 4 \times 10^{43}$~erg~s$^{-1}$. Thus, we estimate that the stars contribute about 20~per~cent to the total radiation field within $\sim3300\,r_\mathrm{sub}$ of the nucleus of NGC 3227. 

\noindent\textit{NG 7469.} For NGC 7469, we can make a similar assessment. \citet{davies2007} report a $K$-band luminosity of the stellar component with the same 0.8~arcsec aperture. Given their approximate conversion to bolometric stellar luminosity, we can estimate $L_\mathrm{NSC}(\mathrm{NGC~7469}) \sim 8 \times 10^{43}$~erg~s$^{-1}$ within $\sim3000\,r_\mathrm{sub}$. Using the X-ray luminosity of the AGN in Table~\ref{tab:lxrsub} and the bolometric conversion by \citet{marconi2004}, we obtain a bolometric luminosity of the AGN of $L_\mathrm{AGN}(\mathrm{NGC~7469}) \sim 3 \times 10^{44}$~erg~s$^{-1}$. From that, we estimate a stellar contribution of $\sim$20~per~cent to the total radiation field.

In summary, from energy conservation we can conclude that the AGN, compared to stars, does indeed have the potential to provide the photons necessary for excitation of the PAHs. There are two caveats to consider in this respect: From AGN unification, we may expect only about half of the solid angle around the AGN to receive significant UV emission necessary to excite the PAHs. Thus, the actual share of AGN contribution may have to be lowered accordingly. On the other hand, the AGN emits much more UV photons than star formation for a given luminosity given the respective spectral shapes. This effect counters the solid angle effect. While it is not possible to rule out star formation as a significant contributor to the PAH emission, the point that needs to be stressed is that the radiation field of the AGN is very strong on these scales and should be considered at least as likely to excite PAHs as stars from an energy perspective. 

We conclude that AGN heating/excitation of PAH-containing dusty gas clouds is a viable option to explain the observed narrow range of PAH surface fluxes and the common radial slopes in our AGN sample with PAH detection below $\sim7\,000$\,\rsub. In reality, we expect that both AGN and NSC contribute to the heating/excitation of the PAHs. Stars will probably dominate at larger distances of $\ga$500\,pc, where the AGN radiation field becomes too small. These are the scales where the well-known PAH emission-star formation relations have been established. 

Our findings do not contradict previously established relations between the 11.3\mumeter PAH feature and star formation rates on larger scales in AGN \citep[e.g.,][]{diamondstanic2010,diamondstanic2012}. Using the 11.3\mumeter PAH feature as a tracer of star formation in AGN relies on two assumptions: 1) that the feature is not suppressed in AGN, and 2) that the [Ne\,{\sc ii}] 12.8\mumeter emission, to which the 11.3\mumeter PAH star formation estimates are calibrated, is a reliable tracer of star formation in AGN. While the 11.3\mumeter PAH feature is less suppressed than features at 6.2, 7.7, and 8.6\mumeter in AGN \citep{smith2007,diamondstanic2010}, it is not clear whether it is not suppressed at all. Also, the [Ne\,{\sc ii}] emission has its own limitations as a star formation tracer in AGN. First, the [Ne\,{\sc ii}] emission can have a significant contribution from AGN emission \citep{groves2006,melendez2008,pereira-santaella2010}, which, if not properly accounted for, will lead to overestimated star formation rates. Indeed, a strong Baldwin effect, commonly associated with AGN excitation, has been reported for the [Ne\,{\sc ii}] \citep{honig2008,keremedjiev2009}. These points, together with the common radial slopes and energetic viability of the AGN to excite PAH features, should raise caution in adopting the 11.3\mumeter PAH feature on small scales in circumnuclear environment of AGN. Indeed, a correlation between AGN luminosity and PAH emission on these small scales may be either interpreted in terms of an AGN-star formation relation or as indicative of PAH excitation by the AGN.

\section{Conclusions}\label{sec:conclusion}

We analyse ground-based mid-IR spectra of the nuclear region of a sample of 28 nearby AGN to study the radial emission profiles of the 11.3\,$\micron$ PAH feature. The data map the PAH emission on unprecedented small spatial scales below 1 kpc. Out of the 28 objects, we detect PAH features in 13 and establish radial profiles of the emitted surface flux for 12 of the sources. To compare the observations on a distance- and luminosity-independent spatial scale, we normalise the emitted surface flux to the dust sublimation radius of each object. As a result, the incident AGN emission at each normalised (=intrinsic) radius is the same in all objects. The sublimation radii are determined based on a new relation between intrinsic X-ray luminosity and sublimation radius. We conclude:

\begin{itemize}
	\item We find a clear radial decrease of the emitted PAH surface flux in the inner $\sim$7,000 sublimation radii ($\sim$500\,pc) around the AGN. The slope of the profile and its normalisation are essentially universal in all objects when normalised by an AGN-intrinsic scale $\propto L^{1/2}$ (here, the sublimation radius). The normalisation of PAH surface fluxes shows a scatter of less than 1 order of magnitude in AGN-intrinsic scales while the PAH luminosities scatter over more than 3 orders of magnitude otherwise.
	\item Circinus is an obvious outlier to this trend, which may be related to strong extinction in the nuclear environment.
    \item By fitting a power-law to the observations of each object, we find a common slope of about $-1.1\pm0.1$ when combining the power-law slopes from the individual objects. 
	\item We argue that a compact emission source is required to explain the common slopes and test the AGN and a nuclear star cluster (NSC) as possible sources of PAH heating/excitation. A simple CLOUDY model of reprocessed AGN radiation is able to reproduce the observed slopes and absolute fluxes. Energy conservation arguments show that the AGN has the potential to provide the radiation field needed to excite PAHs.
	\item About 50~per~cent of the AGN in the total sample have no detection of a 11.3\,$\micron$ PAH feature in our data. About half of these sources have upper limits that fall within or above our detections, meaning that they may need higher S/N data to reveal the features. The rest seem to have genuinely weak features or lack those. 
\end{itemize}

While there exist well-established relations between the PAH emission and star formation on larger scales, our results require some caution when trying to invoke these relations within tens to hundreds of parsecs of an AGN. The James Webb Space Telescope will allow for mapping the radial profiles on similar scales for a much larges sample of AGN and test our findings. Although previous studies have found the 11.3\mumeter PAH feature to be most robust around AGN, we would expect a similar behaviour for other PAH features in the mid-IR, but due to the differential suppression of these features, slopes and normalisations of the radial PAH profiles is expected to differ.

\section*{Acknowledgments}
We thank the anonymous referee for critical and constructive comments that led to significant improvements of the manuscript. The Dark Cosmology Centre is funded by the Danish National Research Foundation. S.F.H. acknowledges support via a Marie Curie International Incoming Fellowship within the 7th EC Framework (PIIF-GA-2013-623804) and the Horizon 2020 ERC Starting Grant \textit{DUST-IN-THE-WIND} (ERC-2015-StG-677117). S.R. thanks Aryabhatta Research Institute of Observational Sciences (ARIES), Nainital, India for providing local support and hospitality during the preparation of this manuscript. A.A.-H. acknowledges financial support from the Spanish Ministry of Economy and Competitiveness through grant AYA2015-64346-C2-1-P. P.G. thanks STFC for support (grant reference ST/J003697/2).

This research has made use of the NASA/IPAC Extragalactic Database (NED) which is operated by JPL, Caltech, under contract with NASA. This research has made use of the VizieR catalogue access tool, CDS, Strasbourg, France. The original description of the VizieR service was published in A\&AS 143, 23.

\bibliographystyle{mnras}
\bibliography{pahagn}

\appendix
\section[]{Spectra of individual objects}\label{app:notes}
\begin{figure*}
	\includegraphics[width=230mm,angle=90]{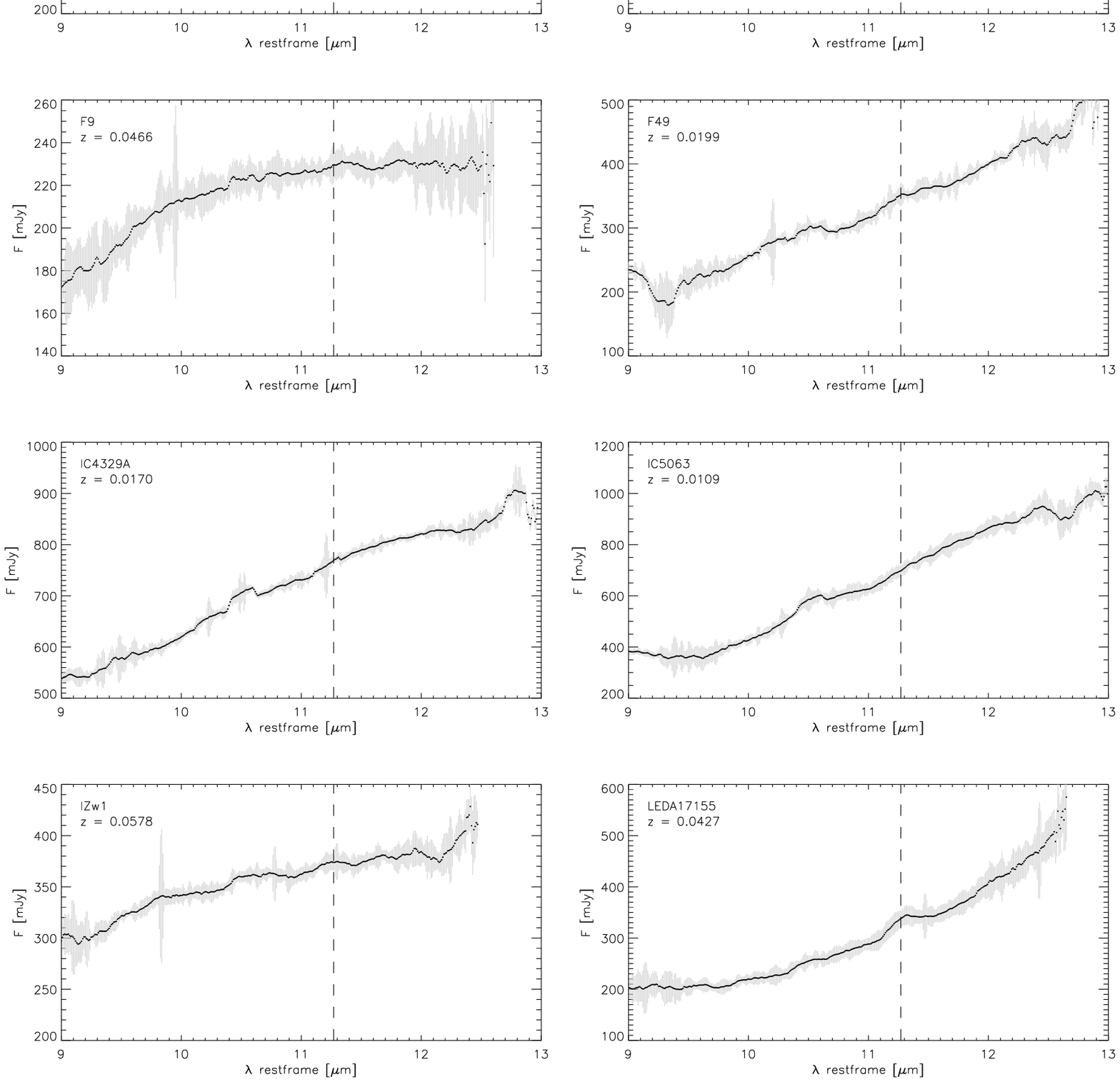}
	\caption{Spectra of the sample extracted within the central aperture with a width of 0.4~arcsec. The position of the 11.3\mumeter PAH feature is indicated by the dashed line in each plot. The data has been smoothed with a 20-pixel boxcar for display purposes.}
	\label{fig:pahspectra1}
\end{figure*}

\begin{figure*}
	\includegraphics[width=230mm,angle=90]{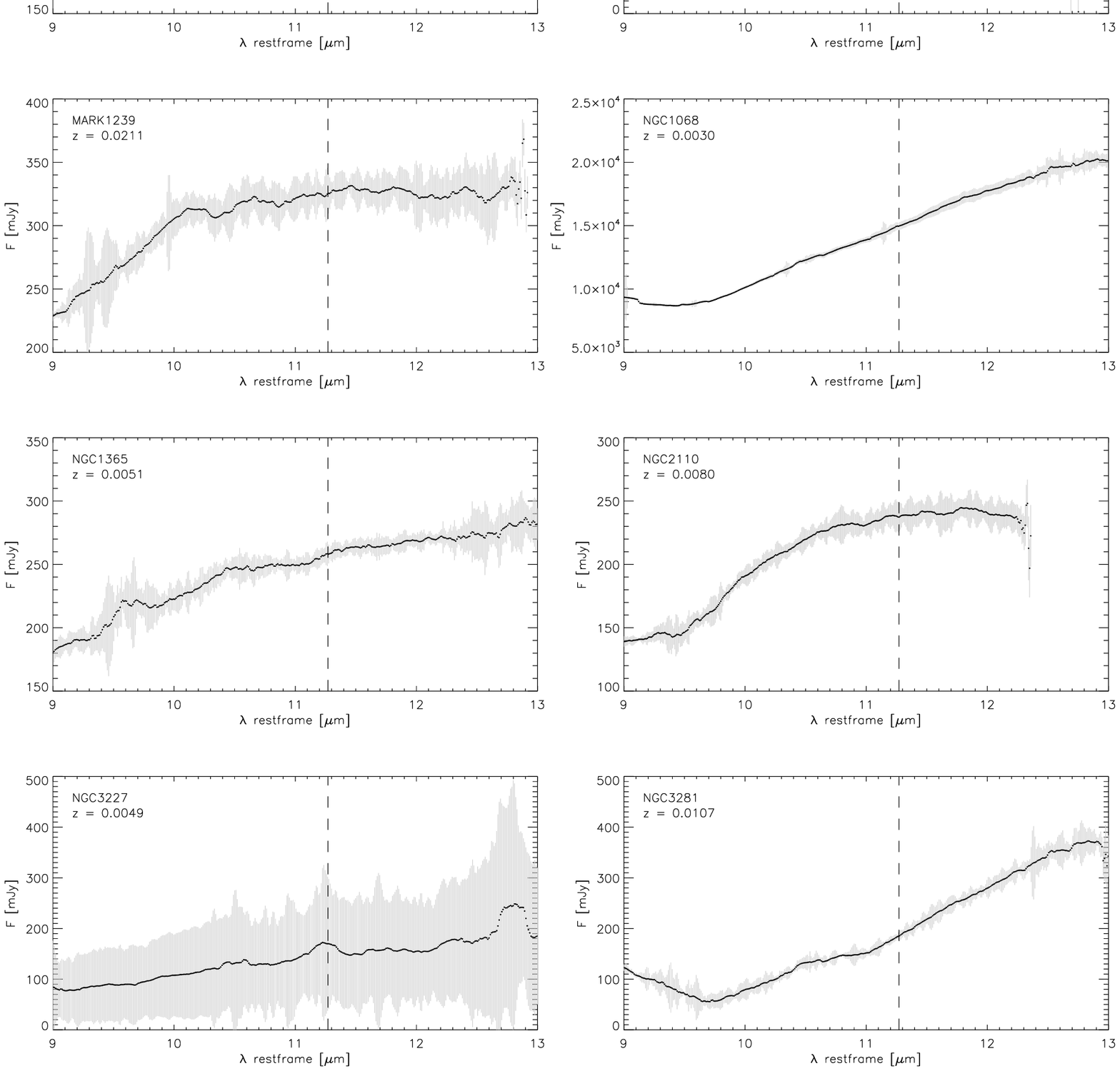}
	\caption{Spectra of the sample extracted within the central aperture with a width of 0.4~arcsec. The position of the 11.3\mumeter PAH feature is indicated by the dashed line in each plot. The data has been smoothed with a 20-pixel boxcar for display purposes.}
	\label{fig:pahspectra2}
\end{figure*}

\begin{figure*}
	\includegraphics[width=230mm,angle=90]{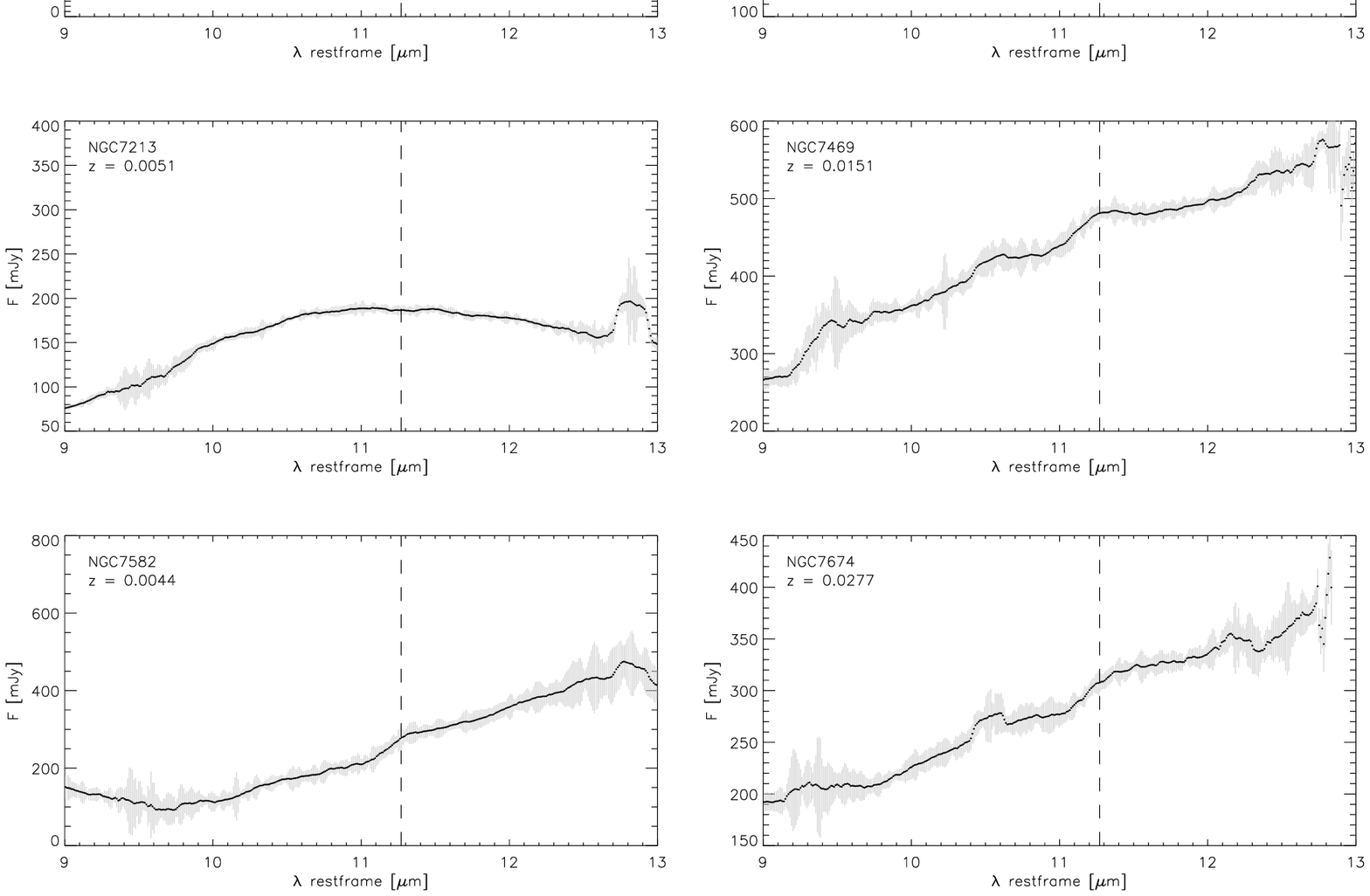}
	\caption{Spectra of the sample extracted within the central aperture with a width of 0.4~arcsec. The position of the 11.3\mumeter PAH feature is indicated by the dashed line in each plot. The data has been smoothed with a 20-pixel boxcar for display purposes.}
	\label{fig:pahspectra3}
\end{figure*}

Figure~\ref{fig:pahspectra1}, Figure~\ref{fig:pahspectra2}, and Figure~\ref{fig:pahspectra3} shows the spectra extracted within the central 0.4~arcsec aperture for all the objects in our sample.


\section[]{Variations in PAH and continuum flux with radius for objects with 11.3\mumeter PAH flux measurements at three or more radii}\label{app:psfcheck}
\begin{figure*}
	\includegraphics[width=105mm,angle=90]{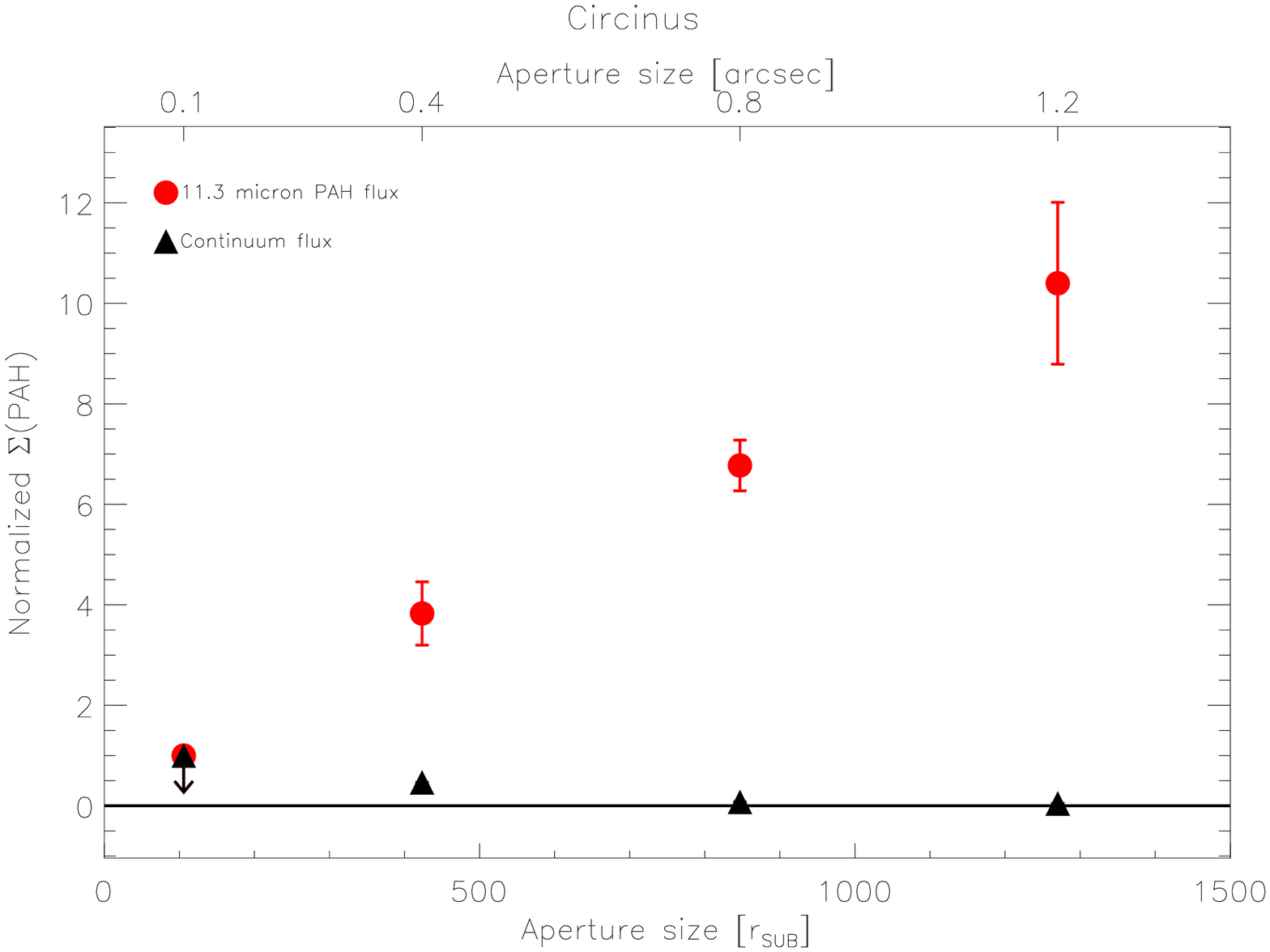}
	\caption{Emitted surface flux of the 11.3\mumeter PAH feature (red circles) and the underlying continuum flux (black triangles) as a function of sublimation radius for Circinus. Arrows denote upper limits.}
	\label{fig:circinus}
\end{figure*}
\begin{figure*}
	\includegraphics[width=105mm,angle=90]{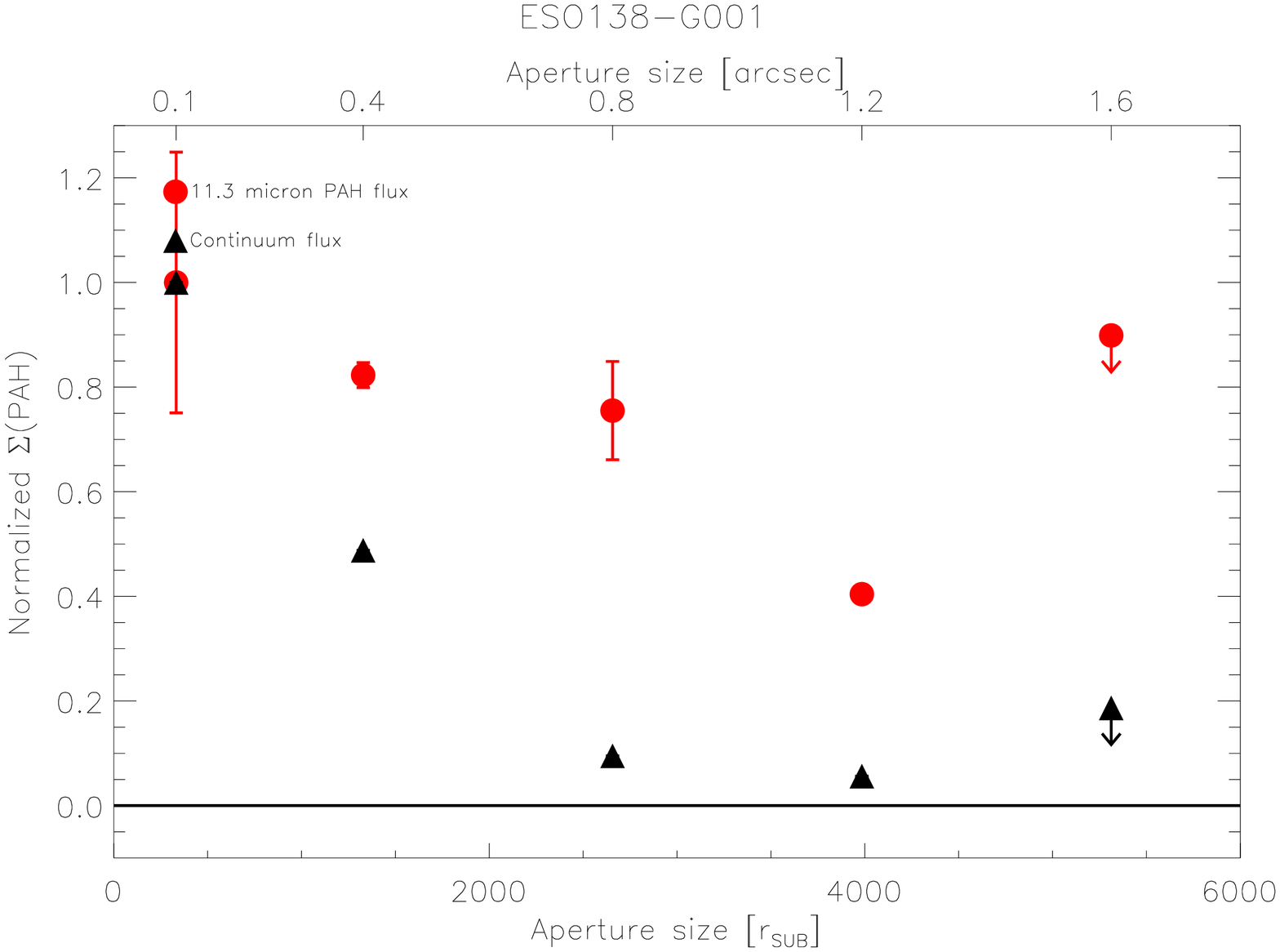}
	\caption{Same as Figure~\ref{fig:circinus} but for ESO138-G001.}
	\label{fig:ESO138-G001}
\end{figure*}
\begin{figure*}
	\includegraphics[width=105mm,angle=90]{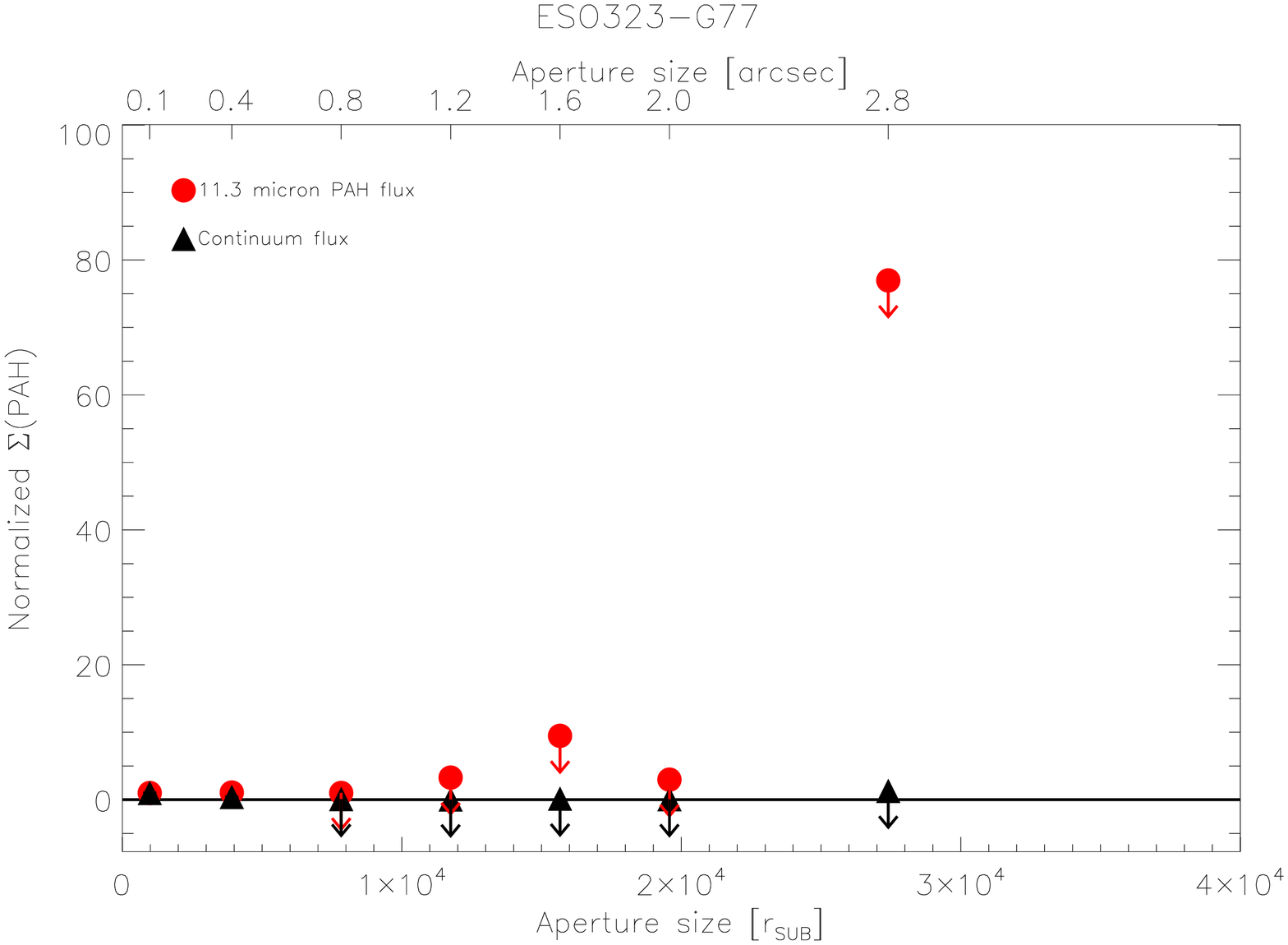}
	\caption{Same as Figure~\ref{fig:circinus} but for ESO323-G77.}
	\label{fig:ESO323-G77}
\end{figure*}
\begin{figure*}
	\includegraphics[width=105mm,angle=90]{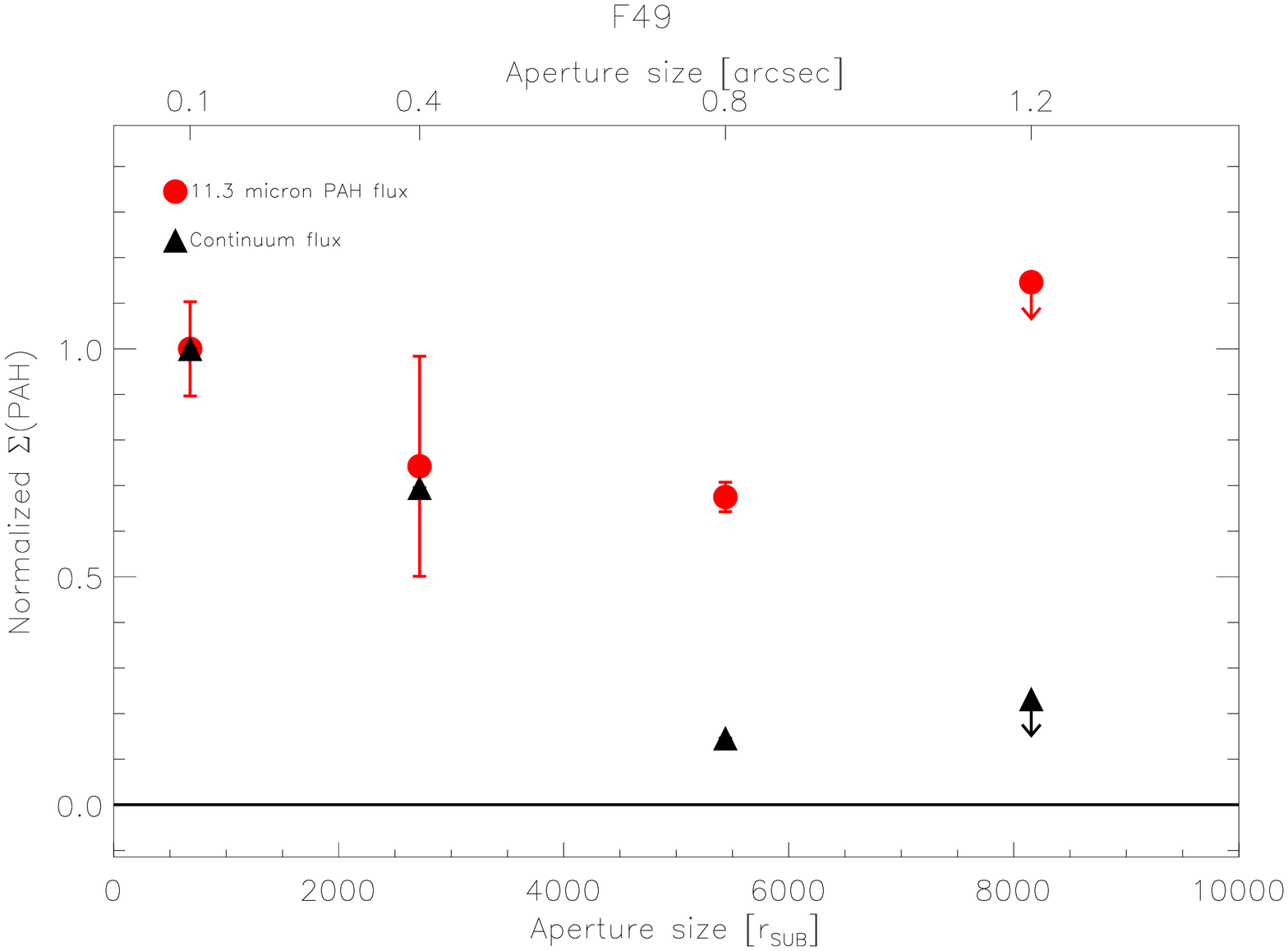}
	\caption{Same as Figure~\ref{fig:circinus} but for F49.}
	\label{fig:F49}
\end{figure*}
\begin{figure*}
	\includegraphics[width=105mm,angle=90]{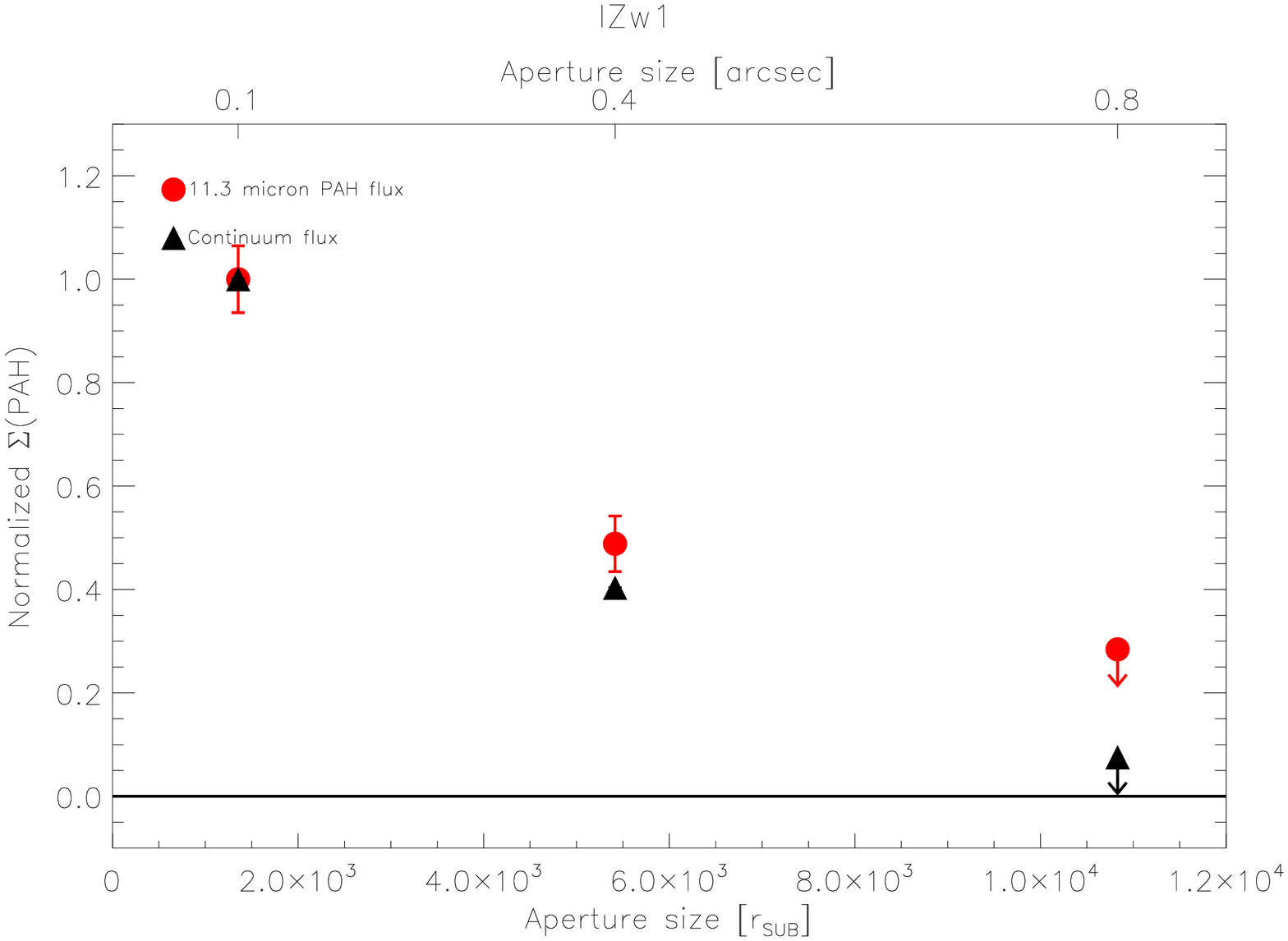}
	\caption{Same as Figure~\ref{fig:circinus} but for IZw1.}
	\label{fig:IZw1}
\end{figure*}
\begin{figure*}
	\includegraphics[width=105mm,angle=90]{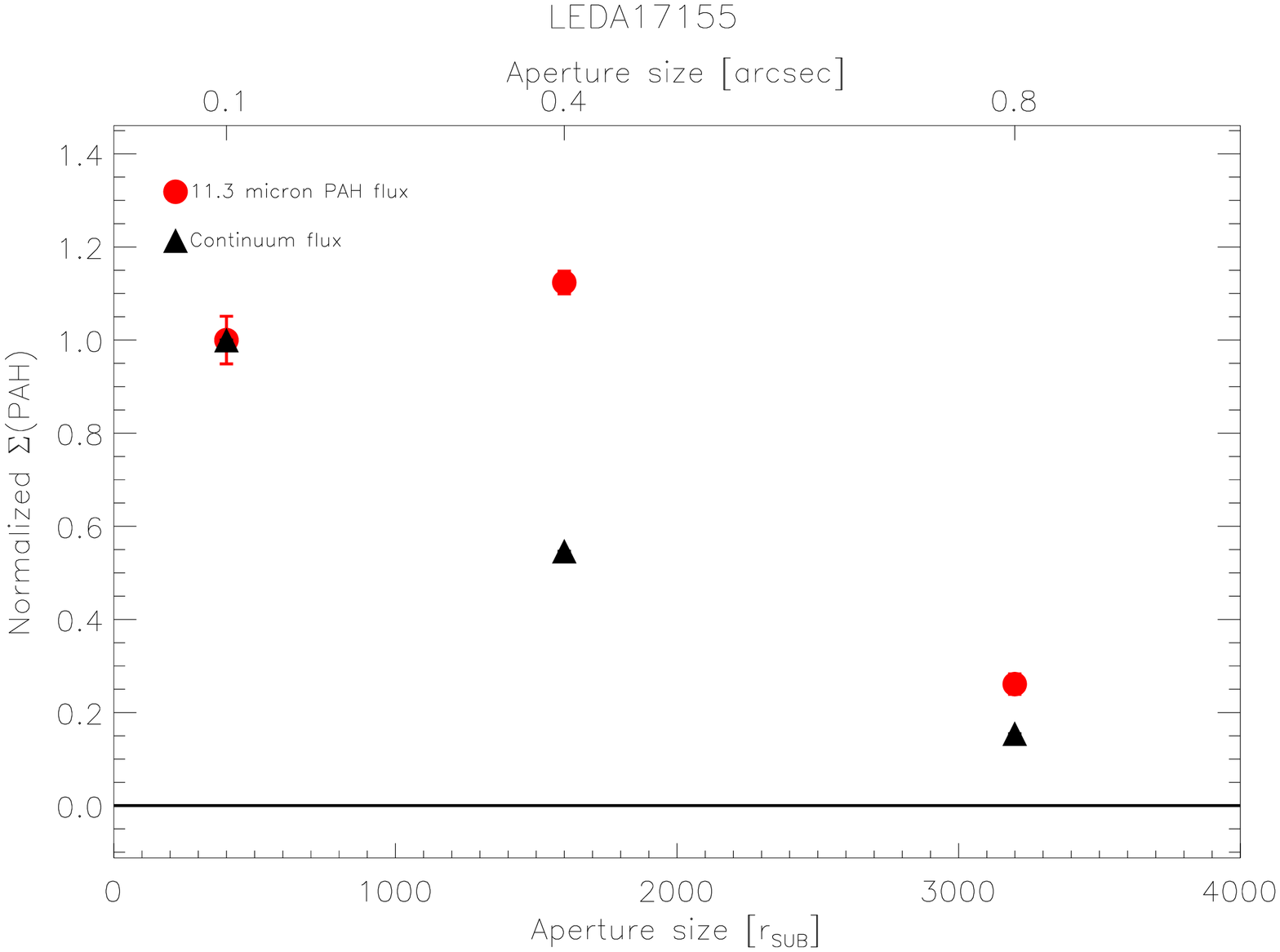}
	\caption{Same as Figure~\ref{fig:circinus} but for LEDA17155.}
	\label{fig:LEDA17155}
\end{figure*}
\begin{figure*}
	\includegraphics[width=105mm,angle=90]{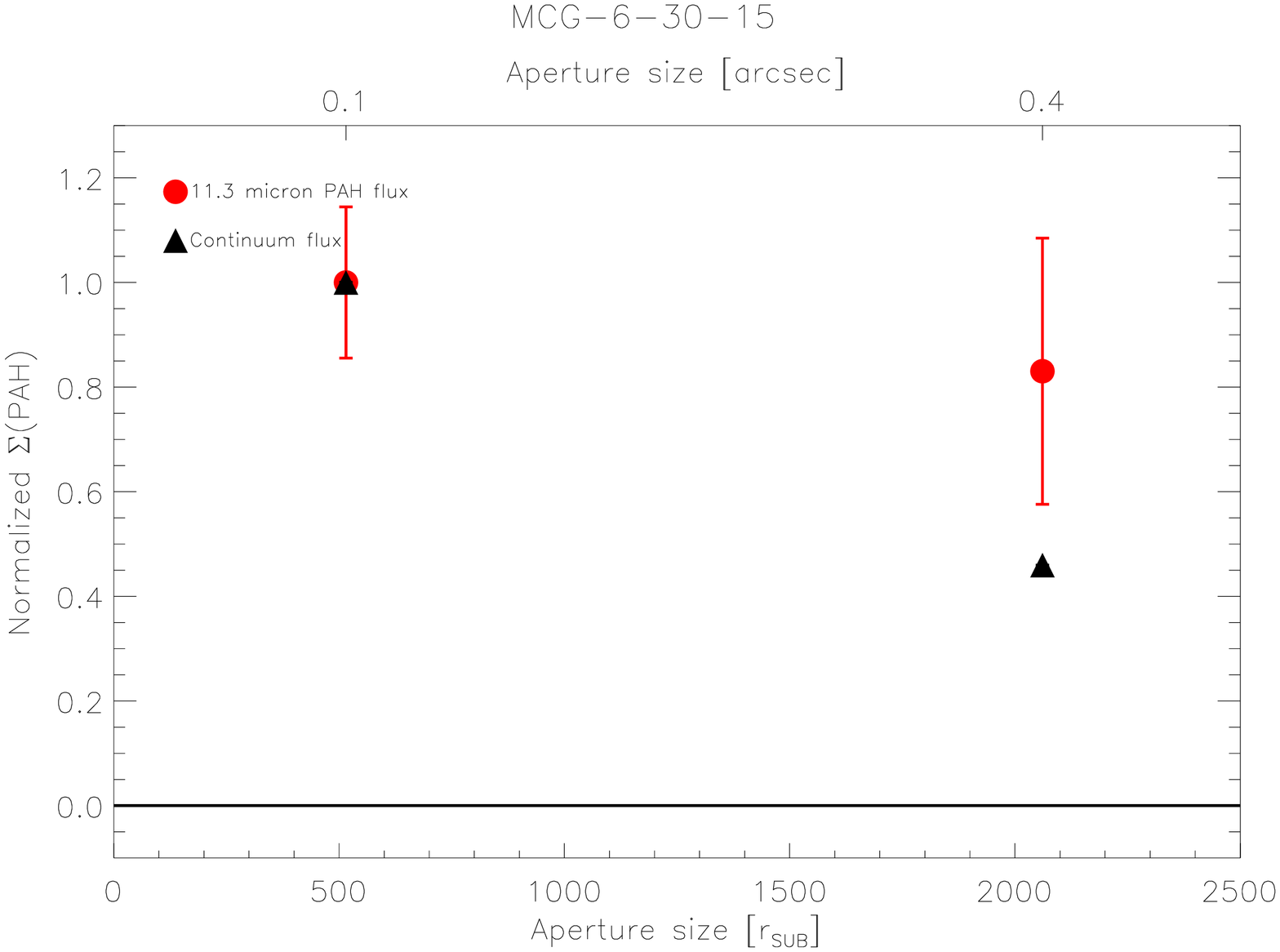}
	\caption{Same as Figure~\ref{fig:circinus} but for MCG-6-30-15.}
	\label{fig:MCG-6-30-15}
\end{figure*}
\begin{figure*}
	\includegraphics[width=105mm,angle=90]{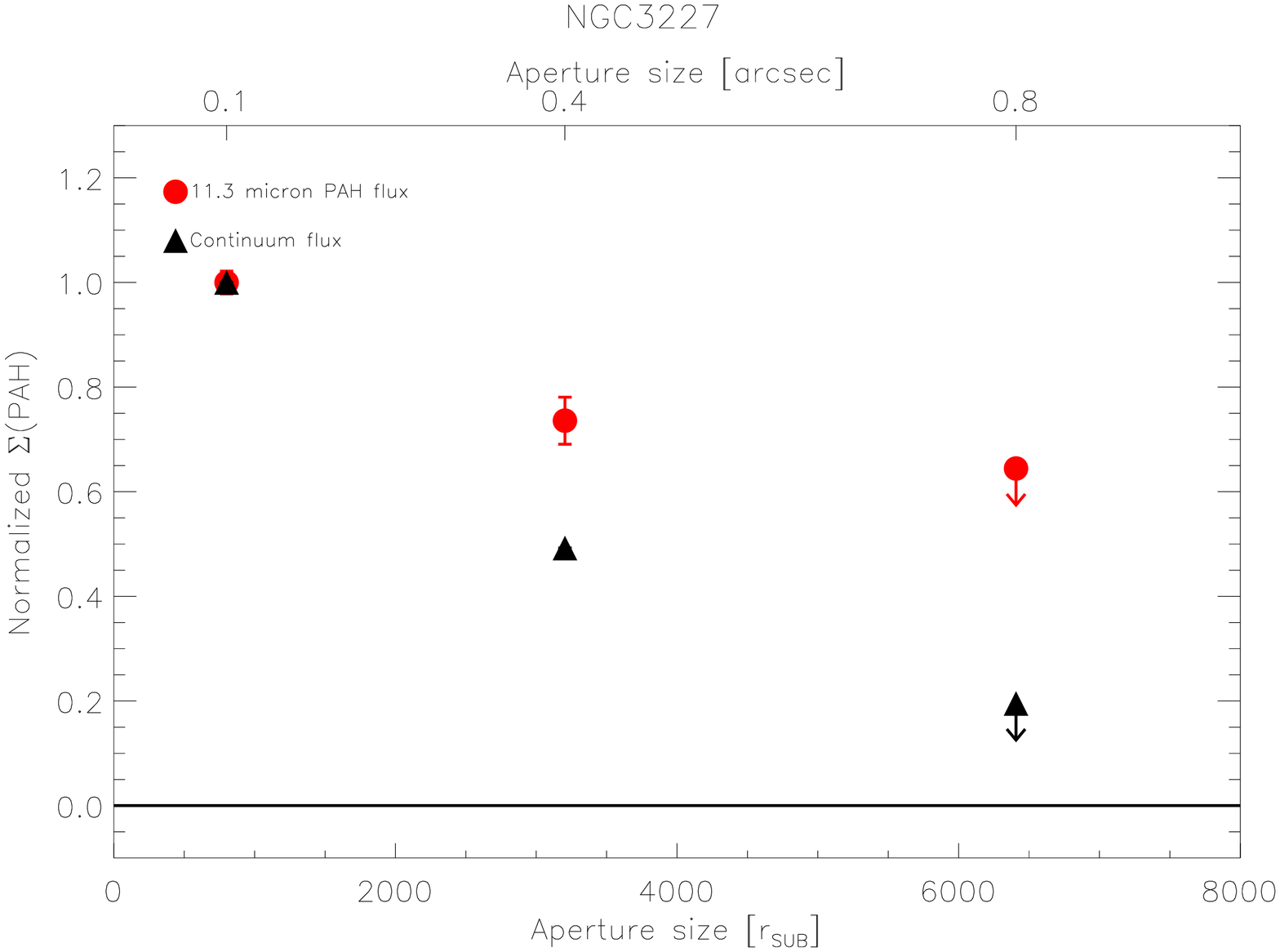}
	\caption{Same as Figure~\ref{fig:circinus} but for NGC3227.}
	\label{fig:NGC3227}
\end{figure*}
\begin{figure*}
	\includegraphics[width=105mm,angle=90]{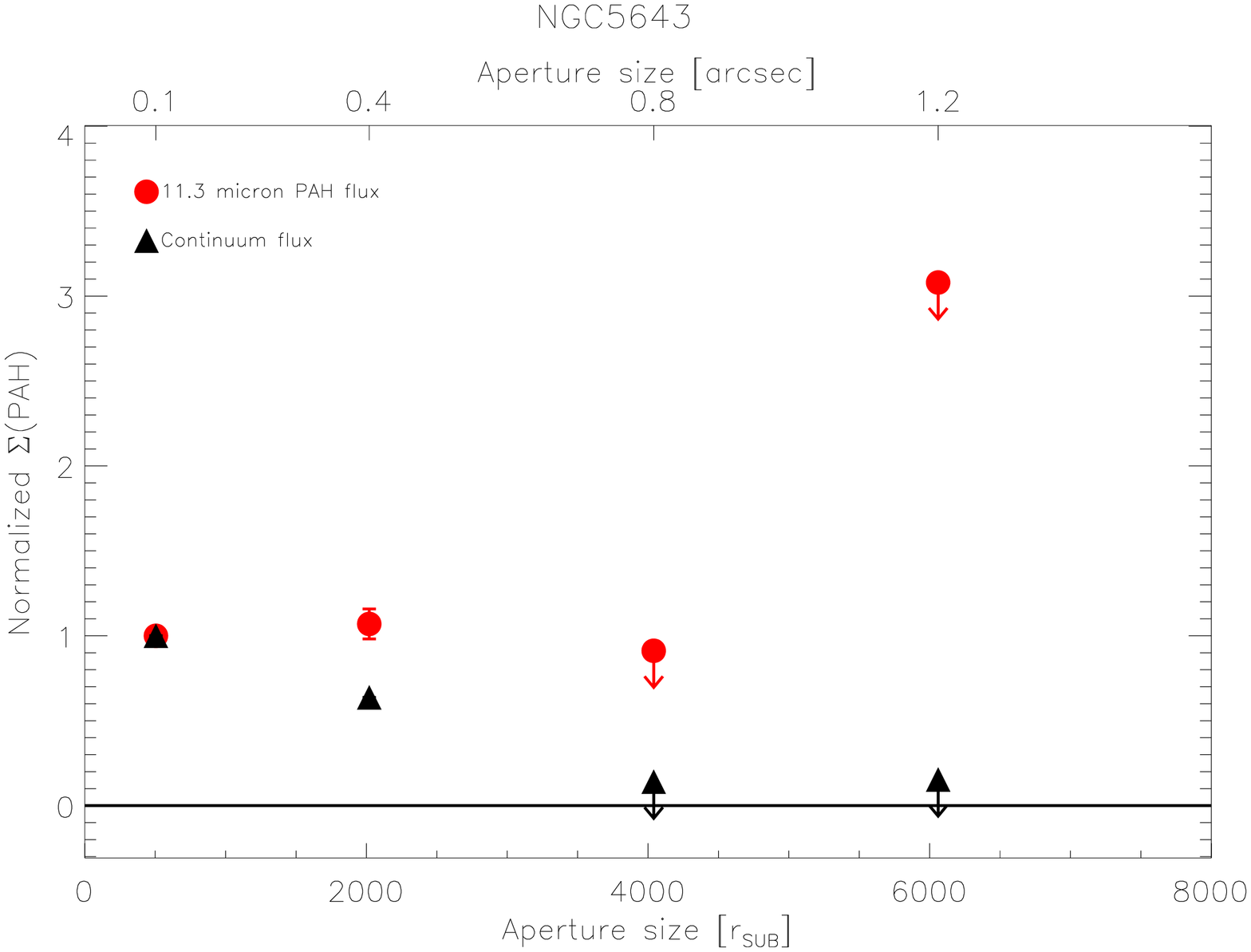}
	\caption{Same as Figure~\ref{fig:circinus} but for NGC5643.}
	\label{fig:NGC5643}
\end{figure*}
\begin{figure*}
	\includegraphics[width=105mm,angle=90]{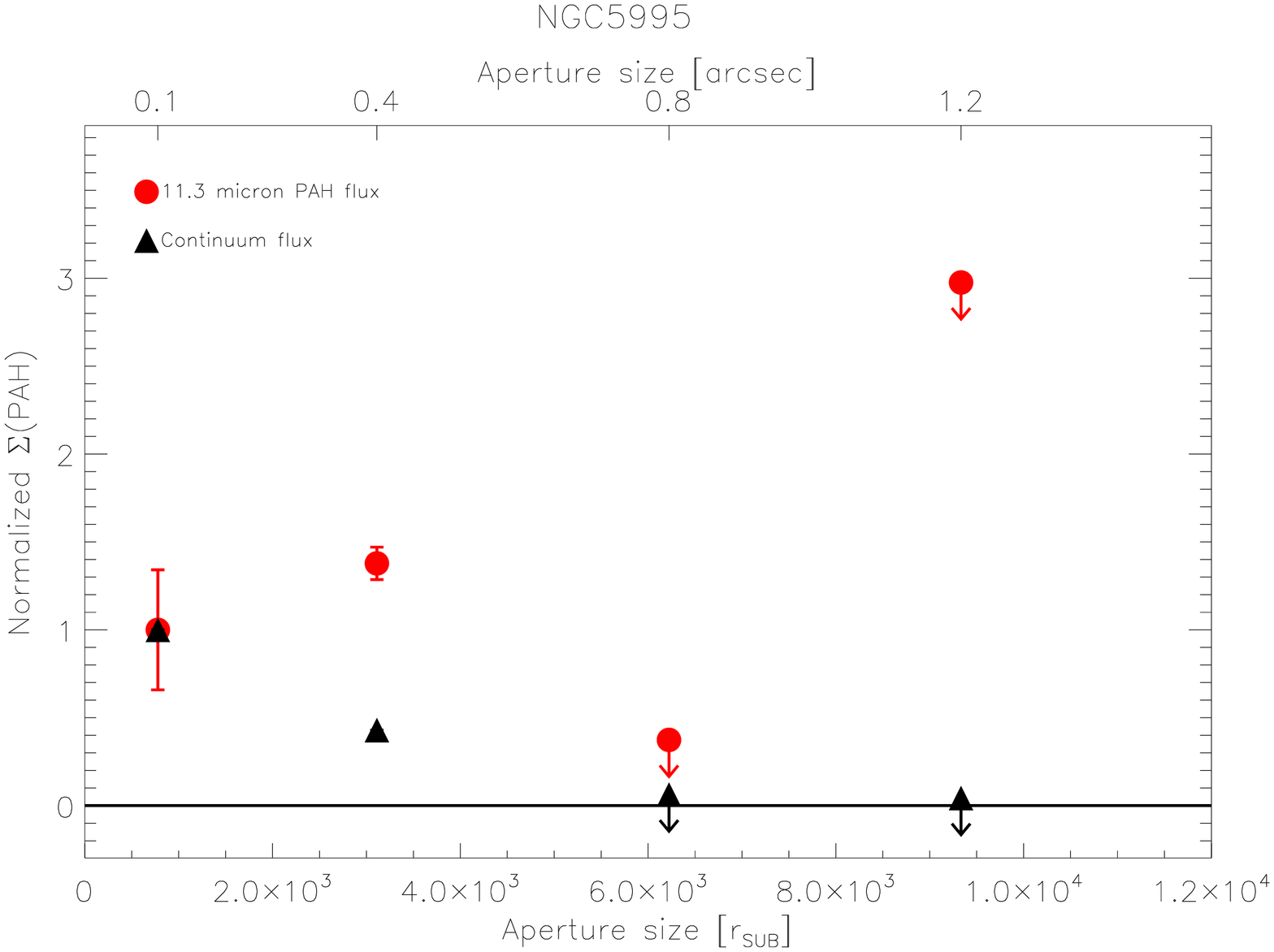}
	\caption{Same as Figure~\ref{fig:circinus} but for NGC5995.}
	\label{fig:NGC5995}
\end{figure*}
\begin{figure*}
	\includegraphics[width=105mm,angle=90]{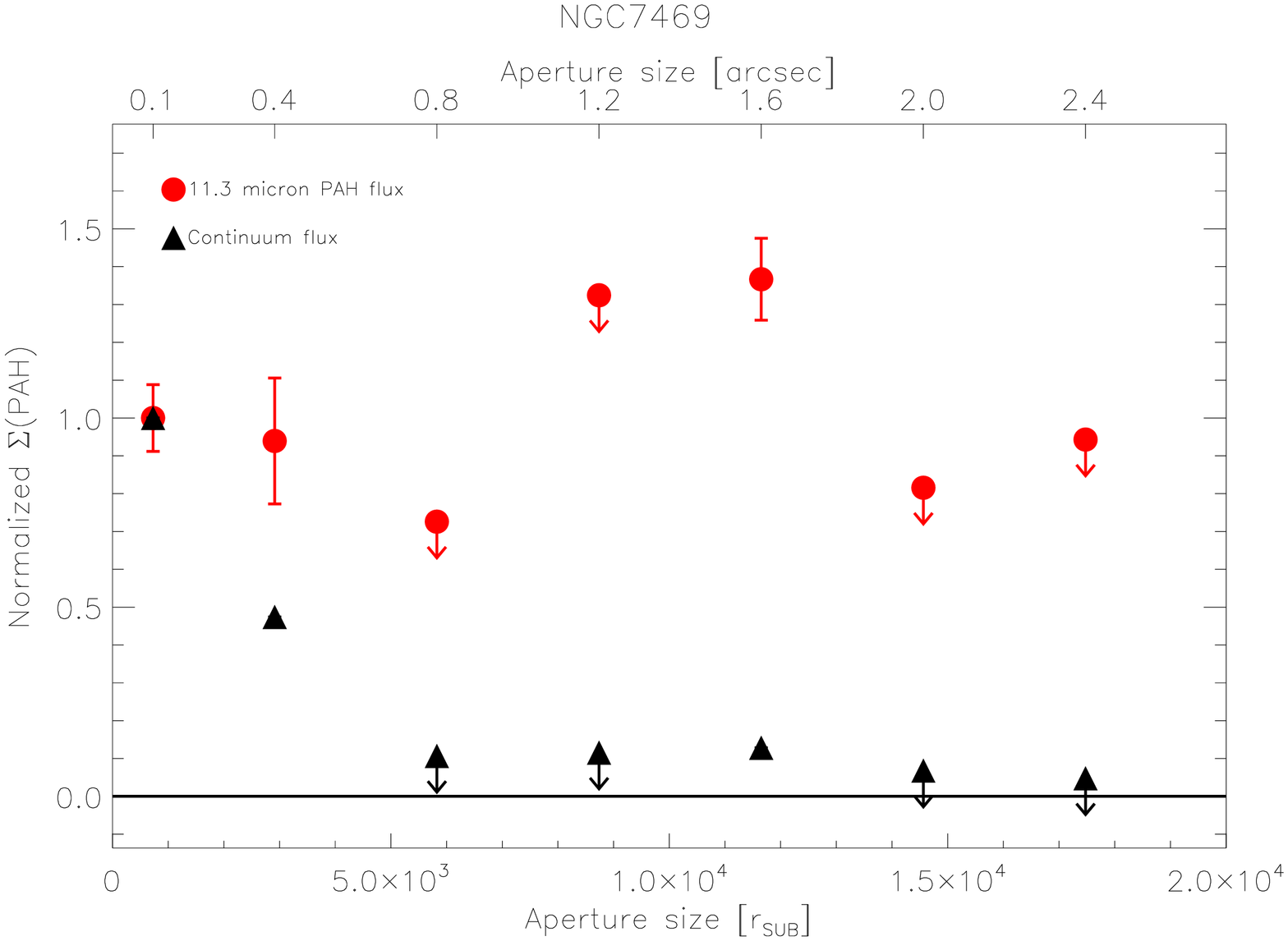}
	\caption{Same as Figure~\ref{fig:circinus} but for NGC7469.}
	\label{fig:NGC7469}
\end{figure*}
\begin{figure*}
	\includegraphics[width=105mm,angle=90]{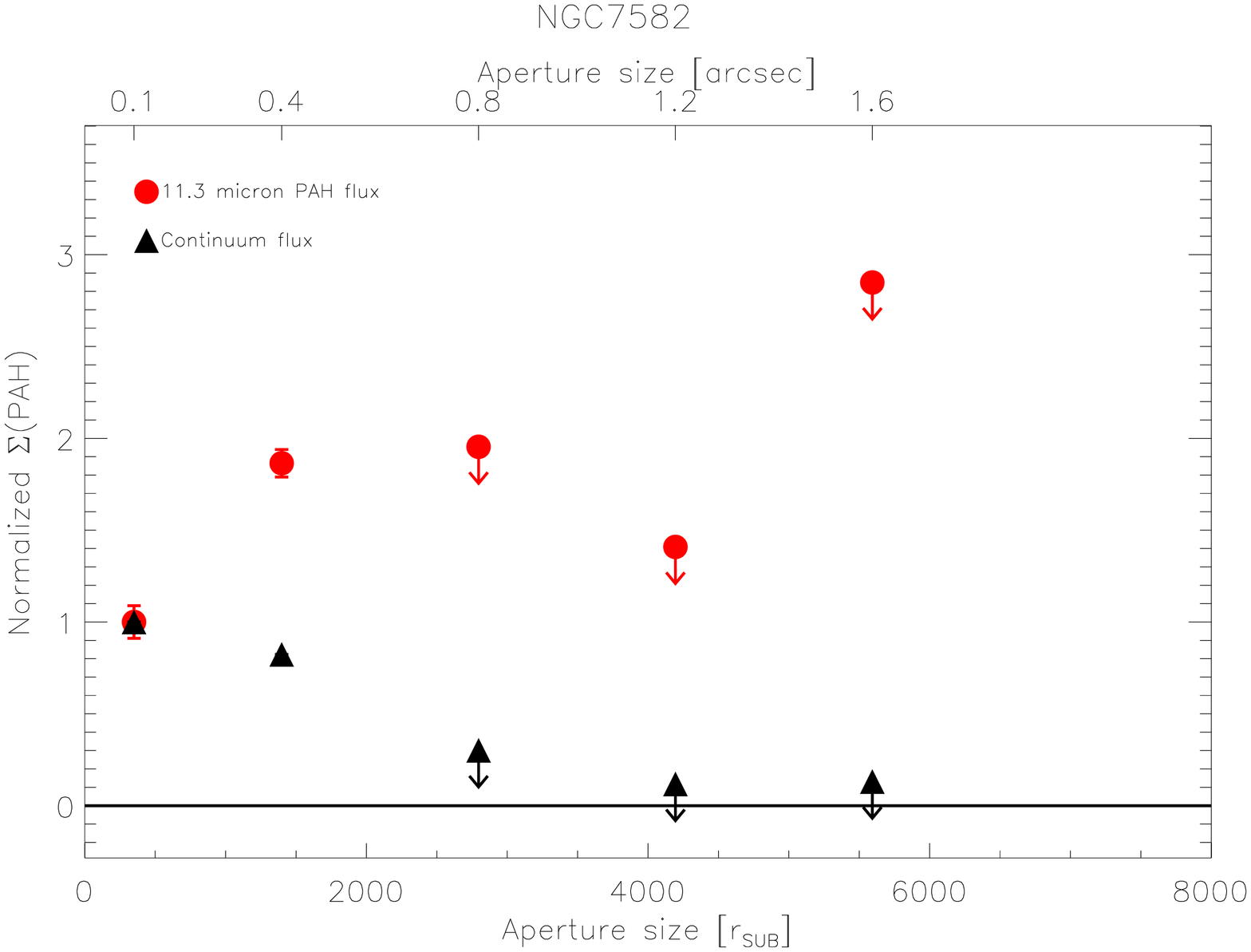}
	\caption{Same as Figure~\ref{fig:circinus} but for NGC7582.}
	\label{fig:NGC7582}
\end{figure*}
\begin{figure*}
	\includegraphics[width=105mm,angle=90]{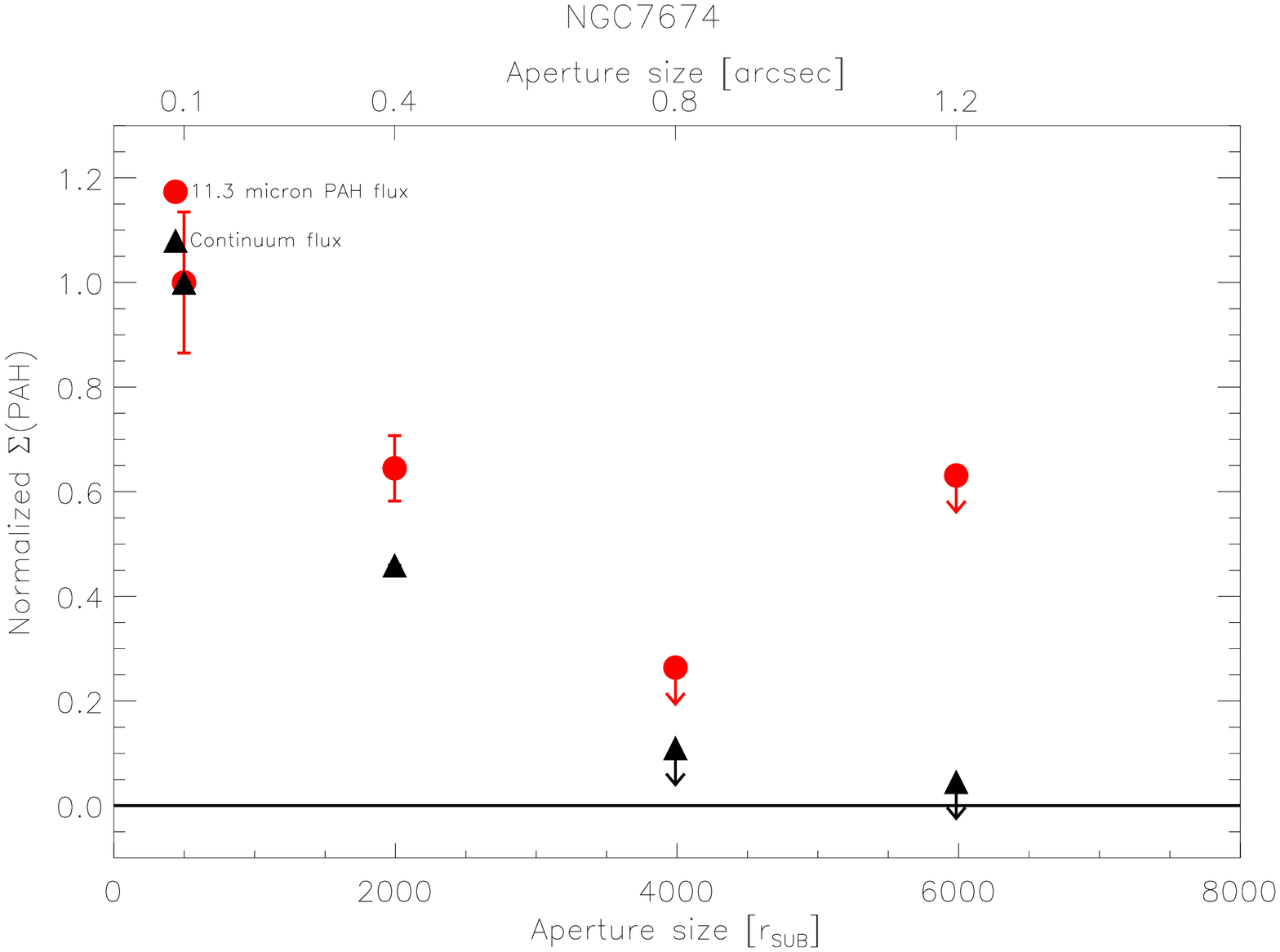}
	\caption{Same as Figure~\ref{fig:circinus} but for NGC7674.}
	\label{fig:NGC7674}
\end{figure*}

\section{Power-law slopes for individual objects and joint probability distribution functions for the PAH slope}\label{app:jpds}

\begin{table*}
	\caption{Slope of the \rsub vs. \pahsurfflux relation for individual objects\label{tab:pahslopeindividual}}
	\begin{tabular}{@{}lcccccc}
		\hline
		\hline
		Object & 3000\rsub & 5000\rsub & 6000\rsub & 7000\rsub & 10000\rsub \\
		\hline
		ESO138-G001	& $-0.86^{+0.28}_{-0.61}$ & $-0.84^{+0.19}_{-0.36}$ & $-0.57^{+0.14}_{-0.25}$ & $-0.58^{+0.15}_{-0.25}$ & $-0.57^{+0.14}_{-0.25}$ \\
		ESO323-G77 	& \nodata & \nodata & \nodata & \nodata & $-0.74^{+0.29}_{-0.62}$ \\
		F49        	& \nodata & \nodata & $-0.91^{+0.27}_{-0.63}$ & $-0.91^{+0.27}_{-0.62}$ & $-0.50^{+0.18}_{-0.35}$ \\
		LEDA17155  	& \nodata & $-1.53^{+0.31}_{-0.73}$ & $-1.52^{+0.30}_{-0.72}$ & $-1.52^{+0.30}_{-0.73}$ & $-1.52^{+0.30}_{-0.74}$ \\
		NGC3227    	& \nodata & \nodata & \nodata & $-0.94^{+0.27}_{-0.63}$ & $-0.93^{+0.27}_{-0.60}$ \\
		NGC5643    	& \nodata & $-0.79^{+0.27}_{-0.62}$ & $-0.80^{+0.28}_{-0.62}$ & $-0.35^{+0.25}_{-0.40}$ & $-0.35^{+0.25}_{-0.41}$ \\
		NGC5995    	& \nodata & \nodata & \nodata & $-1.35^{+0.34}_{-0.72}$ & $-0.55^{+0.29}_{-0.41}$ \\
		NGC7582    	& $-0.41^{+0.30}_{-0.68}$ & $-0.55^{+0.25}_{-0.41}$ & $-0.38^{+0.21}_{-0.33}$ & $-0.39^{+0.21}_{-0.32}$ & $-0.38^{+0.21}_{-0.32}$ \\
		NGC7674    	& \nodata & $-1.40^{+0.28}_{-0.67}$ & $-1.00^{+0.23}_{-0.38}$ & $-1.00^{+0.23}_{-0.38}$ & $-1.00^{+0.23}_{-0.38}$ \\
		NGC7469    	& \nodata & \nodata & $-0.90^{+0.27}_{-0.60}$ & $-0.90^{+0.28}_{-0.62}$ & $-0.66^{+0.25}_{-0.40}$ \\
		\hline
	\end{tabular}	
	\tablecomments{Median power-law slope and $\pm 1 \sigma$ uncertainty for the individual objects at radial cut-offs of 3000, 5000, 6000, 7000, and 10,000 sublimation radii. For a radial cutoff between 5000 and 7000 \rsub, the spearman rank significance for most objects is $<-0.8$.}
\end{table*}

Figures of joint probability distribution functions for the radial PAH power-law slope.

\begin{figure*}
	\includegraphics[width=230mm,angle=90]{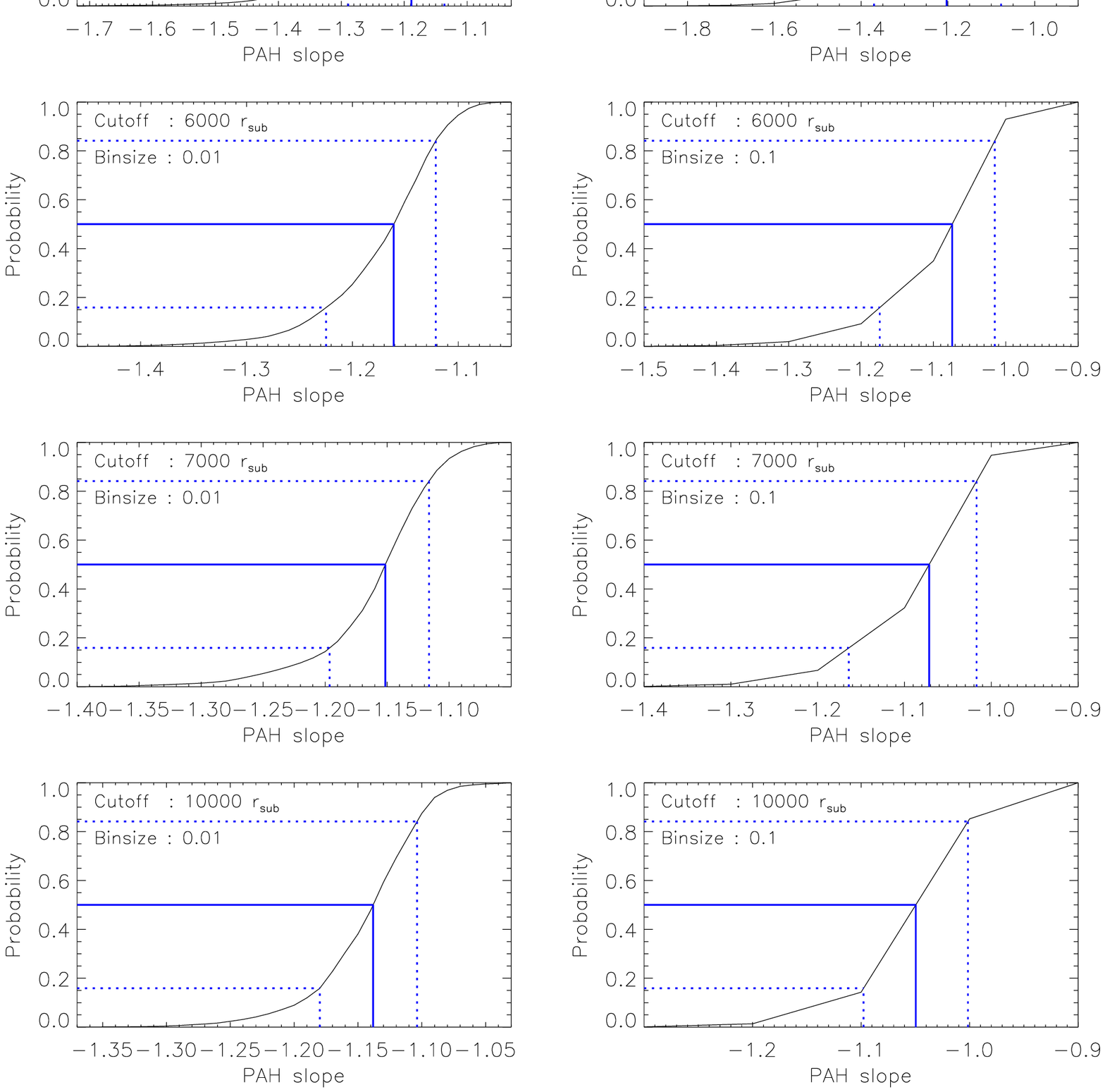}
	\caption{Joint cumulative distribution functions for the PAH power-law slope $\alpha$ for binsizes of 0.01 and 0.1, and radial cut-offs of 3000, 5000, 6000, 7000, and 10,000 sublimation radii. The solid and dashed blue lines represent the median and $\pm 1 \sigma$ uncertainties, respectively.}
	\label{fig:alphajpds}
\end{figure*}

\bsp

\label{lastpage}

\end{document}